\documentclass[
 aip,
 amsmath,amssymb,
 nofootinbib,
reprint
]{revtex4-2}

\usepackage{graphicx}
\usepackage{dcolumn}
\usepackage{bm}

\usepackage[utf8]{inputenc}
\usepackage[T1]{fontenc}
\usepackage{mathptmx}
\usepackage{etoolbox}

\usepackage{float}
\usepackage{comment}

\DeclareMathOperator*{\argmin}{arg\,min}

\usepackage{tikz}
\usetikzlibrary{positioning, arrows.meta}

\usepackage{threeparttable}
\usepackage{caption}
\usepackage{subcaption}

\usepackage{hyperref}
\hypersetup{
    colorlinks=true,
    linkcolor=blue,
    filecolor=magenta,      
    urlcolor=cyan,
    citecolor=blue,
    pdftitle={PINN-1dEBE},
    pdfpagemode=FullScreen,
}

\makeatletter
\def\@email#1#2{%
 \endgroup
 \patchcmd{\titleblock@produce}
  {\frontmatter@RRAPformat}
  {\frontmatter@RRAPformat{\produce@RRAP{*#1\href{mailto:#2}{#2}}}\frontmatter@RRAPformat}
  {}{}
}%
\makeatother
\begin{document}

\preprint{AIP/123-QED}

\title[A physics-informed neural network approach to solve the spatially inhomogeneous electron Boltzmann equation]{A physics-informed neural network approach to solve\\the spatially inhomogeneous electron Boltzmann equation}
\author{Ihda Chaerony Siffa}
\affiliation{Leibniz Institute for Plasma Science and Technology (INP), Felix-Hausdorff-Straße 2, 17489 Greifswald, Germany}%
 \affiliation{Department of Electrical and Information Engineering, Kiel University, Kaiserstraße 2, 24143 Kiel, Germany}
\email{ihda.chaeronysiffa@inp-greifswald.de}
\author{Detlef Loffhagen}%
\author{Markus M. Becker}
\affiliation{Leibniz Institute for Plasma Science and Technology (INP), Felix-Hausdorff-Straße 2, 17489 Greifswald, Germany}%
\author{Jan Trieschmann}
 \affiliation{Department of Electrical and Information Engineering, Kiel University, Kaiserstraße 2, 24143 Kiel, Germany}

\date{\today}

\begin{abstract}
The accurate determination of electron properties is fundamental to low-temperature plasma simulations, necessitating precise solutions to the spatially inhomogeneous electron Boltzmann equation (EBE). This work explores the use of \textit{physics-informed neural networks} (PINNs) for obtaining solutions to the spatially one-dimensional (1D) EBE subject to a uniform electric field in atomic gases. Employing the two-term approximation, the resulting equation for the isotropic distribution is solved directly in kinetic energy space without the conventional transformation to total energy. This approach demonstrates the flexibility of the PINN framework in handling diverse equation formulations. To address the convergence difficulties associated with this class of kinetic equations, a new neural network architecture is introduced. It features a Fourier-feature input layer, adaptive activation functions, and a scaled multiplicative gating mechanism. It is demonstrated that this formulation preserves robust gradient flow throughout the network, which is critical for learning physically correct solutions. Benchmarking against reference data reveals that the present architecture achieves excellent agreement across both microscopic and macroscopic properties of the electrons. Furthermore, the architecture exhibits strong generalization across different gas types and a defined range of electric field strengths without requiring case-specific hyperparameter tuning. Ultimately, the excellent accuracy achieved here validates the applicability of the present PINN method.
\end{abstract}

\maketitle

\section{Introduction}
\label{sec:introduction}
The electron Boltzmann equation (EBE) provides a robust theoretical basis for describing electron kinetics in low-temperature plasmas (LTP)~\cite{golant1980fundamentals, Colonna-2022-ID6107}. Fundamentally, the EBE explicitly resolves the \textit{electron velocity distribution function} (EVDF), which is essential in weakly ionized plasmas, where the EVDF is generally non-Maxwellian~\cite{Godyak-1992-ID3006}. The determination of the EVDF is the prerequisite for obtaining macroscopic quantities such as the electron number density, mean energy, and drift velocity, as well as for calculating reaction rates and transport coefficients. These quantities are of direct physical interest for characterizing electron behavior and, moreover, provide the necessary closure for fluid-Poisson models~\cite{Becker-2013-ID3118}, offering a computationally efficient way to simulate plasmas~\cite{Becker-2017-ID4159, Jovanovic-2023-ID6156}. Ultimately, accurate and efficient simulations of LTP are essential for a vast array of applications, ranging from semiconductor manufacturing to plasma medicine~\cite{Adamovich-2017-ID4238, Adamovich-2022-ID6011}. 

Historically, numerical solutions to the EBE have faced a severe \textit{curse of dimensionality}. While the steady-state equation is six-dimensional in phase space, common practice in LTP restricts the problem to a spatially homogeneous (0D) approximation, solved strictly in energy space via Legendre polynomial expansions~\cite{Hagelaar-2005-ID2276, Leyh-1998-ID1222, Colonna-2022-ID6107}. However, the 0D approximation fails in regimes where the electron energy relaxation length $\lambda_\epsilon$ is comparable to or greater than the characteristic length scale of the system $\Lambda$~\cite{Arslanbekov-1998-ID3840, Tsendin-0000-ID6436,Tsendin-1995-ID976, Winkler-2002-ID1758}. Consequently, the assumption of 0D models that the EVDF is determined by the \textit{local electric field}~\cite{Winkler-1987-ID6437,Kortshagen-1993-ID6438} breaks down. Instead, the system enters a nonlocal kinetic regime where the electron state at (any) given position is influenced by the spatial transport of energy gained from the potential profile (electric field) across neighboring regions~\cite{Sigeneger-1999-ID1309,Kortshagen-1996-ID3034}. To address 1D spatial inhomogeneity, suitable numerical methods were developed in Refs.~\onlinecite{Sigeneger-1995-ID6451, Sigeneger-1996-ID1050, Sigeneger-1998-ID1242, Loffhagen-2002-ID1724, Loffhagen-2005-ID3906, Winkler-1997-ID1137, Loffhagen-2022-ID6521} within the framework of two-term and multi-term approximations, which involve expanding the angular dependence of the EVDF using Legendre polynomials. These methods notably require a transformation of the resulting system of equations for the isotropic and anisotropic distribution from kinetic to total energy space. Despite these advancements, solving the EBE remains a challenging endeavor, both mathematically and numerically. This has prompted recent investigations into Scientific Machine Learning (SciML) as a potential alternative approach for tackling this class of kinetic equations~\cite{Kawaguchi-2020-ID6439, Kawaguchi-2022-ID6158, Kim-2023-ID6440}.

SciML is rapidly emerging as a powerful field aiming to accelerate scientific discovery and computational tasks through the integration of machine learning (ML) methods \cite{Baker-2019-ID6443, Jumper-2021-ID6444, Cranmer2020, Anirudh-2022-ID6250}. Within computational physics and engineering, a notable SciML development is \textit{physics-informed neural networks}~(PINNs) \cite{Raissi-2019-ID6187}, which offer a new approach to numerically solving partial differential equations (PDEs) in both forward and inverse problem settings with great flexibility. Unlike conventional discretization methods, PINNs represent a solution to a PDE as a neural network, which is a continuous and differentiable function, and optimize its parameters to satisfy the governing PDE and associated boundary and initial conditions~\cite{Lagaris-1998-ID6186}. This approach has shown promise for various applications, particularly for high-dimensional PDEs~\cite{Sirignano-2018-ID6446, Hu-2024-ID6447} and reconciling governing equations with experimental data, e.g.\ ill-posed inverse problems~\cite{Raissi-2019-ID6187,Raissi-2020-ID6452}. PINNs often operate in an unsupervised or self-supervised manner for forward problems by relying solely on the equation and boundary-initial conditions~\cite{Lu-2021-ID6189}, analogous to conventional numerical solvers.

Nonetheless, PINNs face notable challenges. Studies have shown that PINNs can struggle with complex PDEs, sometimes converging to non-physical solutions or exhibiting sub-optimal convergence rates compared to the well-established numerical techniques, thereby necessitating ongoing research into robust neural network architectures and optimization strategies~\cite{Krishnapriyan2021, Wang-2021-ID6448}. It is worth noting that the principles underpinning PINNs, such as incorporating physical constraints or governing equations, can facilitate the development of data-efficient machine-learned surrogate solvers~\cite{Wang-2021-ID6196,Li-2024-ID6449} for ML-accelerated physical simulations~\cite{Gergs-2023-ID6462, Trieschmann-2023-ID6461, Huang-2025-ID6450}.

Recent works have demonstrated the applicability of PINNs to the spatially homogeneous EBE directly in velocity space~\cite{Kawaguchi-2020-ID6439, Kawaguchi-2022-ID6158}, and in doing so, eliminating the need for Legendre polynomial expansion of the EVDF. To bring this approach closer to practice, the work in Ref.~\onlinecite{Kim-2023-ID6440} has recently proposed a parametric model for the homogeneous case. At the same time, the spatial dependence remains unaddressed.

{\sloppy
This work presents the first data-free PINN approach for solving the 1D spatially inhomogeneous electron Boltzmann equation subject to a uniform electric field under the two-term approximation framework. The present approach circumvents the traditional total-energy transformation, solving the system directly in position and kinetic energy space. A gating-based neural network architecture with a Fourier-feature layer is introduced that mitigates convergence failures faced by the recent architectures. The method is validated against reference solutions for noble gases such as neon, argon, krypton, and xenon across a defined range of electric field strengths, demonstrating the accuracy and applicability of this PINN approach.}

\section{Physical background}
\label{sec:problem_setup}
\subsection{Two-term approximated electron Boltzmann equation}
\label{sec:twoterm_ebe}
The present work studies the spatial evolution of electrons in weakly-ionized, non-isothermal plasmas governed by the steady-state electron Boltzmann equation
\begin{equation}
\label{eq:full_ebe}
    \boldsymbol{v} \cdot \nabla_{\boldsymbol{x}} F - \frac{e_0}{m_\mathrm{e}} \boldsymbol{E} \cdot \nabla_{\boldsymbol{v}}F = C^\mathrm{el}\left[F\right] + \sum_{k\in \mathcal{P}} C^\mathrm{in}_{k}\left[F\right]. 
\end{equation}
Here, $F(\boldsymbol{x},\boldsymbol{v})$ denotes the EVDF in six-dimensional phase space ($\boldsymbol{x}, \boldsymbol{v}$). The quantity $\boldsymbol{E}$ represents the electric field, $e_0$ is the elementary charge, and $m_\mathrm{e}$ is the electron mass. The collision operators on the right-hand side account for elastic ($C^\mathrm{el}$) and various ($k$) inelastic ($C^\mathrm{in}_k$) processes between electrons and heavy particles.

Directly solving Eq.~\eqref{eq:full_ebe} presents significant computational and implementation challenges due to its high dimensionality and the complex collision integrals. Consequently, standard practice in technological plasma modeling involves reducing the dimensionality under appropriate physical assumptions. To this end, a spatially inhomogeneous plasma with spatial gradients and the electric field aligned along the $z$-direction, i.e. $\boldsymbol{E}=E(z)\hat{e}_z$, is considered. Under this condition, the system exhibits symmetrical behavior around the electric field axis, which allows reducing the dependence of the EVDF to $F(z,v,v_z/v)$ on the spatial coordinate $z$, velocity magnitude $v=|\boldsymbol{v}|$, and direction cosine $v_z/v$ of the velocity.

Now, the EVDF can be expressed in terms of kinetic energy $U=m_\mathrm{e}v^2 / 2$ and direction cosine $v_z/v = \cos \theta$, where $\theta$ is the polar angle in velocity space relative to the electric field. By expanding the EVDF in a series of Legendre polynomials $P_l(\cos \theta)$ and truncating after the first two terms ($P_0(\cos \theta)=1$ and $P_1(\cos \theta)=v_z/v$), the two-term representation is given by~\cite{Loffhagen-2005-ID3906}
\begin{equation}
    \label{eq:evdf_twoterm}
    F(z,U,v_z/v) = \frac{1}{2 \pi} \left(\frac{m_\mathrm{e}}{2}\right)^{3/2} \left(f_0(z,U) + f_1(z,U)\frac{v_z}{v}\right),
\end{equation}
where $f_0(z,U)$ and $f_1(z,U)$ are the isotropic and anisotropic part of the EVDF, respectively. Substituting Eq.~\eqref{eq:evdf_twoterm} into equation Eq.~\eqref{eq:full_ebe} yields the following system of PDEs for $f_0$ and $f_1$:
\begin{widetext}
\begin{equation}
\label{eq:f_0}
\begin{aligned}
    \frac{1}{3} U \frac{\partial f_1 }{\partial z} - \frac{e_0}{3} E(z) \frac{\partial}{\partial U} \left(U f_1\right) - \frac{\partial}{\partial U} \left[ 2\frac{m_\mathrm{e}}{M} U^2 N Q^\mathrm{el}(U) \left(f_0 + k_\mathrm{B} T \frac{\partial}{\partial U} f_0\right) \right] \\
    + \sum_{k \in \mathcal{P}} U N Q_k^\mathrm{in}(U) f_0 = \sum_{k \in \mathcal{P}} (U + U_k^\mathrm{in} ) N Q_k^\mathrm{in}(U + U_k^\mathrm{in}) f_0(z, U + U_k^\mathrm{in}), 
\end{aligned}
\end{equation}
\begin{equation}
\label{eq:f_1}
\begin{aligned}
    f_1(z,U) = \frac{1}{N Q^\Sigma(U)}\left(-\frac{\partial f_0}{\partial z} + e_0E(z)\frac{\partial f_0}{\partial U} \right).
\end{aligned}
\end{equation}
\end{widetext}
Here, $N$ is the density of the background gas with mass $M$ and temperature $T$, and $k_\mathrm{B}$ is the Boltzmann constant. The functions $Q^\mathrm{el}(U)$ and $Q^\mathrm{in}_k(U)$ denote the energy-dependent cross sections for momentum transfer in elastic collisions and inelastic collision processes (e.g.\ excitation and ionization processes) from a set of inelastic processes $\mathcal{P}$, respectively. $U_k^\mathrm{in}$ is the threshold energy for a $k$-type inelastic collision process, and the total cross section is denoted by $Q^\Sigma(U) = Q^\mathrm{el}(U) + \sum_{k  \in \mathcal{P}} Q_{k}^\text{in}(U)$. For simplicity, the ionization process is treated as an excitation-like process that results in energy loss but does not produce a new electron. To obtain the explicit collision operators, the original collision integrals have been expanded with respect to the mass ratio $m_\mathrm{e}/M$, retaining only the leading term of each expansion.  

Substituting Eq.~\eqref{eq:f_1} into Eq.~\eqref{eq:f_0} and applying the product rule yields the following second-order elliptic equation
\begin{widetext}
\begin{equation}
\begin{aligned}
\label{eq:final_equation}
    &\frac{U}{3} \left[-\frac{Q_\Sigma(U)}{N} \frac{\partial^2 f_0}{\partial z^2} + \frac{Q_\Sigma(U)}{N} \left( \frac{\mathrm{d} (e_0E(z))}{\mathrm{d} z} \frac{\partial f_0}{\partial U} + e_0E(z) \frac{\partial^2 f_0}{\partial z \partial U}\right) \right] - \frac{1}{3} e_0 E(z) \left[-\frac{Q_\Sigma(U)}{N}\frac{\partial f_0}{\partial z} \right. \\ 
    +& Q_\Sigma(U)\frac{e_0 E(z)}{N}\frac{\partial f_0}{\partial U} 
    - \left. \frac{U}{N} \frac{\mathrm{d} Q_\Sigma(U)}{\mathrm{d} U} \frac{\partial f_0}{\partial z} -  \frac{U}{N} Q_\Sigma(U)  \frac{\partial^2 f_0}{\partial z \partial U} + \frac{U}{N} e_0 E(z)\frac{\mathrm{d} Q_\Sigma(U)}{\mathrm{d} U} \frac{\partial f_0}{\partial U} +  \frac{U}{N} e_0 E(z) Q_\Sigma(U) \frac{\partial^2 f_0}{ \partial U^2}\right] \\
    -& 2\frac{m_\mathrm{e}}{M} N\left[2U Q^\mathrm{el}(U)f_0 + U^2 \frac{\mathrm{d} Q^\mathrm{el}(U)}{\mathrm{d} U}f_0 + U^2 Q^\mathrm{el}(U) \frac{\partial f_0}{\partial U} + k_\mathrm{B} T \left(2 U Q^\mathrm{el}(U)\frac{\partial f_0}{\partial U} +  U^2 \frac{\mathrm{d} Q^\mathrm{el}(U)}{\mathrm{d} U}\frac{\partial f_0}{\partial U} \right. \right. \\
    +& \left. \left. U^2 Q^\mathrm{el}(U) \frac{\partial^2 f_0}{\partial U^2}\right)\right] + \sum_{k \in \mathcal{P}} U N Q_k^\mathrm{in}(U) f_0 = \sum_{k \in \mathcal{P}} (U + U_k^\mathrm{in} ) N Q_k^\mathrm{in}(U + U_k^\mathrm{in}) f_0(z, U + U_k^\mathrm{in}), 
\end{aligned}
\end{equation}
\end{widetext}
where $Q_\Sigma (U) = 1 /Q^\Sigma (U)$ denotes the reciprocal of the total cross section. Note that Eq.~\eqref{eq:final_equation} degenerates as $U\rightarrow0$ (see Appendix~\ref{appendix:degenerate_elliptic}).

In previous works~\cite{Sigeneger-1995-ID6451, Sigeneger-1996-ID1050}, solving the system of Eqs.~\eqref{eq:f_0} and \eqref{eq:f_1} necessitated a coordinate transformation. By substituting the kinetic energy $U$ with the total energy $\epsilon = U + W(z)$, where \mbox{$W(z)=e_0\int_0^z E(\hat z) d\hat z$} is the potential energy in the electric field, the original system was transformed into a single PDE for the isotropic part~\cite{Loffhagen-2005-ID3906}. Recasting the problem into total energy space was a necessary step; it inherently simplifies the mathematical structure by eliminating the mixed derivatives and convective electric field terms appearing in Eq.~\eqref{eq:final_equation}. This allowed the equation to be treated as a more tractable initial boundary value problem, enabling development of suitable numerical solution methods. In contrast, the present work solves the PDE for $f_0(z,U)$ represented in Eq.\eqref{eq:final_equation} directly in the original coordinate system $(z,U)$.

\subsection{Solution domain and boundary conditions}
\label{sec:solution_domain_bc}
Solving Eq.~\eqref{eq:final_equation} requires the definition of the solution domain alongside its corresponding boundary conditions. Let us consider a plasma system bounded spatially by a cathode at $z=0$ and an anode at $z=z_\mathrm{max}$. In kinetic energy space, the domain ranges from zero kinetic energy up to a sufficient cutoff $U_\mathrm{max}$, defining the solution domain $\Omega = [0,z_\mathrm{max}] \times [0, U_\mathrm{max}]$.

\begin{figure}[hb]
\centering
\includegraphics[width=0.48\textwidth]{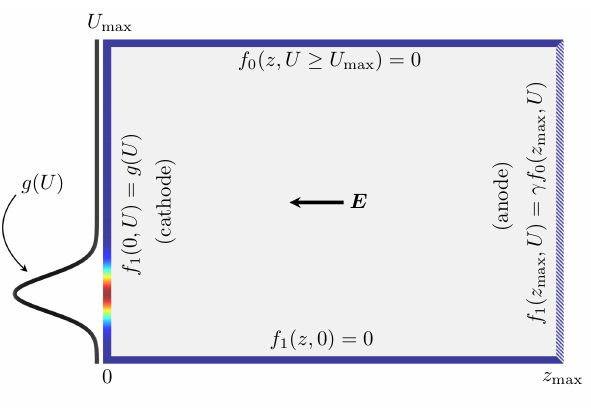}
\caption{\label{fig:solution_domain_bc}Solution domain and boundary conditions.}
\end{figure}

The system is driven by a source generating an influx of electrons at the cathode. These electrons are accelerated towards the anode by an applied electric field, where they are partially reflected back into the domain. Two fundamental physical constraints apply to the energy boundaries. First, at zero kinetic energy ($U=0$), electron velocity is zero thus cannot exhibit directional shift; this implies that the anisotropic part of the EVDF must vanish, $f_1(z,0)=0$. Second, since the probability of an electron having infinitely high energy is effectively zero, it is assumed that at a sufficiently high kinetic energy ($U\geq U_\mathrm{max}$), $f_0 (z, U\geq U_\mathrm{max})=0$ holds for the isotropic part of the EVDF. These physical assumptions form the basis for defining a well-posed set of boundary conditions essential for solving Eq.~\eqref{eq:final_equation}.

The specific boundary conditions and the solution domain are illustrated in Fig.~\ref{fig:solution_domain_bc}. At the cathode $(z=0)$, the electron source is modeled using a prescribed analytical Gaussian distribution, given by
\begin{equation}
\label{eq:g(U)}
\begin{aligned}
    g(U) = \frac{j}{e_0} \frac{h(U)}{(2/m_\mathrm{e})^{1/2} \int_0^{U_\mathrm{max}} U h(U)\,\mathrm{d}U /3}
\end{aligned}
\end{equation}
where the shape function $h(U)$ is defined as
\begin{equation}
\label{eq:h(U)}
\begin{aligned}
    h(U) = U \,\, \mathrm{exp} \left[-\left(\frac{U-U_\mathrm{center}}{U_\mathrm{width}}\right)^2\right].
\end{aligned}
\end{equation}
This condition serves as an influx of electrons into the domain; thus, it is suitable for the anisotropic distribution $f_1(0,U)$ at this boundary. The parameters are the electron current density $j$, the center energy $U_\mathrm{center}$, and the energy width $U_\mathrm{width}$. Together, they determine the amplitude, the width, and the location of the distribution along the $U$-axis. At zero kinetic energy or the \textit{bottom} of the solution domain ($U=0$), the anisotropic part is set to zero as discussed previously. It is important to note that a boundary condition on the isotropic component $f_0(z,0)$ cannot be imposed, as the population density of zero-energy electrons is part of the unknown solution. At the \textit{top} of the solution domain, \mbox{$f_0(z,U\geq U_\mathrm{max})=0$} is specified. Finally, at the anode \mbox{($z=z_\mathrm{max}$)}, electrons are assumed to be partially reflected with a reflection factor $\gamma$. This is expressed as a mixed boundary condition, coupling the two distribution components expressed by \mbox{$f_1(z_\mathrm{max},U) = \gamma f_0(z_\mathrm{max},U)$}. 

\subsection{Macroscopic properties of electrons}
The two-term approximation framework detailed in Sec.~\ref{sec:twoterm_ebe} provides a high-fidelity kinetic description of electrons under specific physical constraints. From this, the spatially dependent macroscopic properties of the electrons can be rigorously derived. These macroscopic quantities are obtained by taking moments of the isotropic and anisotropic distribution component, $f_0(z,U)$ and $f_1(z,U)$. For example, the electron density $n_\mathrm{e}(z)$ and mean energy $\langle U \rangle(z)$ are obtained by integrating the isotropic part $f_0 (z,U)$ over the kinetic energy $U$ as follows
\begin{equation}
\label{eq:electron_density}
n_\mathrm{e}(z) = \int_0^\infty f_0(z,U) \sqrt{U} \,\mathrm{d}U,
\end{equation}
\begin{equation}
\label{eq:mean_energy}
\langle U \rangle (z) = \frac{1}{n_\mathrm{e}(z)}\int_0^\infty Uf_0(z,U) \sqrt{U}  \,\mathrm{d}U.
\end{equation}
The directional transport is characterized by the anisotropic component $f_1 (z,U)$. Specifically, the macroscopic electron particle flux $j_z(z)$ and electron energy flux $j_{\mathrm{e}z}(z)$ are obtained by taking the moments of $f_1(z,U)$ as
\begin{equation}
\label{eq:particle_flux}
j_z(z) = \frac{1}{3}\sqrt{\frac{2}{m_\mathrm{e}}} \int_0^\infty U f_1(z, U)\,\mathrm{d}U,
\end{equation}
\begin{equation}
\label{eq:energy_flux}
j_{\mathrm{e}z}(z) = \frac{1}{3}\sqrt{\frac{2}{m_\mathrm{e}}} \int_0^\infty U^2 f_1(z, U)\,\mathrm{d}U.
\end{equation}
The transport properties such as mobility $b(z)$ and diffusion coefficient $D(z)$ are given by
\begin{equation}
\label{eq:mobility}
b(z) = -\frac{e_0}{3} \sqrt{\frac{2}{m_\mathrm{e}}} \int_0^\infty \frac{U}{N Q^\Sigma (U)} \frac{\partial}{\partial U} \left(\frac{f_0(z,U)}{n_\mathrm{e}(z)} \right)\mathrm{d}U,
\end{equation}
\begin{equation}
\label{eq:diffusion_coeff}
D(z) = \frac{1}{3} \sqrt{\frac{2}{m_\mathrm{e}}} \int_0^\infty \frac{U}{N Q^\Sigma (U)} \frac{f_0(z,U)}{n_\mathrm{e}(z)}\,\mathrm{d}U,
\end{equation}
and the rate coefficient for a $k$-type inelastic collision process is calculated as
\begin{equation}
\label{eq:rate_coeff}
k_k(z) = \frac{1}{n_\mathrm{e} (z)} \sqrt{\frac{2}{m_\mathrm{e}}} \int_0^\infty U Q_k^{\mathrm{in}}(U) f_0 (z,U) \, \mathrm{d}U.
\end{equation}

It is well established from experiments and kinetic studies that these properties are nonlocal in conditions when $\lambda_\epsilon \geq \Lambda$, meaning the electron energy relaxation length equals or exceeds the characteristic system scale. Even within regions where the electric field is spatially uniform, as investigated in this work, electrons undergo a complex spatial relaxation process. They are accelerated by the field until inelastic threshold energies are reached. Subsequently, these electrons may lose energy via inelastic collisions and return to a lower energy state before being accelerated again. This cycle creates spatial structure that appears as striations, and a highly non-Maxwellian distribution that varies spatially~\cite{raizer1991gas, Tsendin_2010}.

\section{PINN architecture to solve the electron Boltzmann equation}
\subsection{Brief overview of PINNs}
\label{sec:pinn_overview}
The PINN approach seeks to find an approximate solution to a PDE by incorporating the governing physics directly in the network's training process. This is typically achieved by \textit{softly} constraining the network to satisfy the differential equation and its boundary and initial conditions. Despite the fact that this method can also incorporate sparse measurement data, the present work examines the use of PINNs as a \textit{neural solver} for a forward problem where no measurement data is present.

Let us consider a general steady-state partial differential equation with variable coefficients in the strong form
\begin{align}
    \label{eq:residual_pde}
    \mathcal{D}\left[u,\mathcal{Q}\right](\boldsymbol{x})&= S(\boldsymbol{x}), ~~~ \forall \boldsymbol{x} \in \Omega \\
    \label{eq:boundary_pde}
    \mathcal{B}\left[u,\mathcal{Q}\right](\boldsymbol{x})&= G(\boldsymbol{x}), ~~~ \forall \boldsymbol{x} \in \partial\Omega
\end{align}
where $u(\boldsymbol{x})$ is the function of interest (the unknown solution) within a domain $\Omega \subset \mathbb{R}^d$ in $d$ dimensions. $\mathcal{D}$ and $\mathcal{B}$ are the \textit{differential} and \textit{boundary operators}, respectively, which may include derivatives of high order as well as mixed derivatives. The set of variable coefficients is represented by $\mathcal{Q}$, the source term by $S(\boldsymbol{x})$, and $G(\boldsymbol{x})$ prescribes the required behavior at the boundary. For a simple Dirichlet condition, the operator $\mathcal{B}$ would simply evaluate the function $u(\boldsymbol{x})$. Suppose that there is a neural network $u_\theta(\boldsymbol{x})$ parameterized by trainable weights and biases $\theta$ that can approximate the target function $u(\boldsymbol{x})$, i.e.~a network with sufficient complexity. By substituting $u(\boldsymbol{x})$ with $u_\theta(\boldsymbol{x})$ in Eqs.~\eqref{eq:residual_pde} and \eqref{eq:boundary_pde}, a composite loss function with $M$ objectives for this physics-informed neural network can be constructed. This function takes the form of a weighted sum
\begin{equation}
\label{eq:general_loss}
    \mathcal{L}(\theta) = w_\mathrm{f}\mathcal{L}_\mathrm{f}(\theta)+\sum_{j=1}^{M-1} w_j \mathcal{L}_{j}(\theta),
\end{equation}
where $\mathcal{L}_\mathrm{f}(\theta)$ represents the PDE residual $L_p$-loss term given by
\begin{equation}
\label{eq:pde_loss}
    \mathcal{L}_\mathrm{f}(\theta) = \frac{1}{n_\mathrm{f}} \sum_{i=1}^{n_\mathrm{f}} \|\mathcal{D}\left[u_\theta,\mathcal{Q}\right](\boldsymbol{x}_\mathrm{f}^{(i)}) - S(\boldsymbol{x}_\mathrm{f}^{(i)})\|_p,
\end{equation}
and $\mathcal{L}_{j}(\theta)$ denotes the $j$-th boundary $L_p$-loss term defined as
\begin{equation}
\label{eq:boundary_loss}
    \mathcal{L}_{j}(\theta) = \frac{1}{n_{j}} \sum_{i=1}^{n_{j}} \|\mathcal{B}\left[u_\theta,\mathcal{Q}\right](\boldsymbol{x}_j^{(i)}) - G_j(\boldsymbol{x}_j^{(i)})\|_p.
\end{equation}
The hyperparameters $w_\mathrm{(\cdot)}$ and $n_\mathrm{(\cdot)}$ are the tunable weight and the number of collocation points for each respective loss term. Importantly in the PINN approach, the derivatives required to evaluate $\mathcal{D}[u_\theta]$ and $\mathcal{B}[u_\theta]$ are computed using automatic differentiation~\cite{Baydin2017}.

{\sloppy
Consequently, the target function $u(\boldsymbol{x})$ can be learned by minimizing Eq.~\eqref{eq:general_loss}, i.e.~optimizing $u_\theta(\boldsymbol{x})$ following the learning objective $\theta^* = \argmin_{\theta} \mathcal{L}(\theta)$ to obtain a set of optimal parameters $\theta^*$ via gradient-based optimization. Following the Universal Approximation Theorem~\cite{Hornik-1989-ID6197, Hornik-1991-ID6198}, a sufficiently wide or deep network can approximate the function to arbitrary accuracy, provided the optimization algorithm can successfully navigate the highly non-convex loss landscape and avoid poor local minima.
}

\subsection{Neural network architectures with gating mechanisms}
\label{sec:network_architecture}
PINNs provide a flexible framework for solving PDEs; however, their training can face significant challenges, especially for complex and stiff PDEs. A common failure mode is poor convergence, where the optimization process gets stuck in a poor local minimum or converges to a non-physical or trivial solution. In contrast to recent works that have focused on improving optimization and sampling strategies~\cite{Krishnapriyan2021, Wang-2021-ID6448,Daw2023}, the present work focuses on architecture design, specifically a class of deep, fully connected neural networks with gating mechanisms. 

This neural network is denoted by $u_\theta (X)$, where the input $X\in \mathbb{R}^{n \times d}$ is a matrix consisting of $n$ collocation points and $d$ physical dimensions. The network evaluates to an output $Y=u_\theta (X)$ through the following forward pass
\begin{subequations}
\begin{alignat}{2}
    &\mathcal{F}(X) = \begin{bmatrix} 
    \sin(XB)\\
    \cos(XB)
    \end{bmatrix}, &\quad& B \sim \mathcal{N}(\mu=0,\sigma^2) \\
    &H^{(0)} = \mathcal{F}(X), \\
    &K = \phi\left(a^{(1)}H^{(0)}W^K + b^K\right), \\
    &V = \phi\left(a^{(1)}H^{(0)}W^V + b^V\right), \\
    &Z^{(l)} = H^{(l-1)}W^{(l)} + b^{(l)}, &\quad& l=1,...,L \\
    &H^{(l)} = \mathrm{Gat}\left(a^{(l+1)}, Z^{(l)}, K, V\right), &\quad& l=1,...,L \\
    &H^{(L+1)} = H^{(L)}W^{(L+1)} + b^{(L+1)}, \\
    &Y = \exp\left(-H^{(L+1)}\right).
\end{alignat}
\end{subequations}

The sequence begins by mapping the input $X$ into a higher dimension representation via the Fourier-feature mapping function $\mathcal{F}(X)$, resulting in the initial state $H^{(0)} \in \mathbb{R}^{n \times m}$. The mapping uses a non-trainable random matrix $B \in \mathbb{R}^{d \times \frac{m}{2}}$~\cite{Tancik2020}, where $m$ represents the number of neurons in each hidden layer. The weights in $B$ are sampled from a zero-mean normal distribution with standard deviation $\sigma$. $W^{(\cdot)}$ and $b^{(\cdot)}$ represent the trainable weights and biases, respectively. The dimensions of $W^{(\cdot)}$ depend on the specific layer: $\mathbb{R}^{d\times m}$ for the initial transformation, $\mathbb{R}^{m\times m}$ for the hidden layers, and $\mathbb{R}^{m\times 1}$ for the final layer $W^{(L+1)}$ to map the last hidden state to a scalar output. To improve the network's convergence properties, adaptive activation functions are employed~\cite{Jagtap2020}. Accordingly, a set of trainable scalar parameters $a^{(l)}$ is introduced, which control the slope of the activation functions at layer $l$.

\begin{figure*}[ht]
    \centering
    \resizebox{0.8\textwidth}{!}{%
        \begin{tikzpicture}[
            >={Stealth[length=2.5mm, width=1.5mm]},
            box/.style={draw, very thick, minimum width=1.5cm, minimum height=2cm, align=center},
            txt/.style={fill=white, inner sep=2pt}
        ]

        \node[box] (zU) {$X$};
        \node[box, right=1.5cm of zU] (H0) {$H^{(0)}$};
        \node[box, right=1.5cm of H0] (H1) {$H^{(1)}$};
        \node[right=1.5cm of H1] (dots) {$\cdots$};
        \node[box, right=1.5cm of dots] (HL) {$H^{(L+1)}$};
        \node[box, right=1.5cm of HL] (f0) {$Y$};

        \node[box, above=0.5cm of H0] (K) {$K$};
        \node[box, below=0.5cm of H0] (V) {$V$};

        \draw[->, thick] (zU) -- node[txt] {$\mathcal{F}$} (H0);

        \draw[->, thick] (H0) -- node[txt, pos=0.5] (Gat1) {Gat} (H1);
        \draw[->, thick] (H1) -- node[txt] (Gat2) {Gat} (dots);
        \draw[->, thick] (dots) -- node[txt] (Gat3) {Gat} (HL);

        \draw[->, thick] (HL) -- node[txt] {exp} (f0);

        \draw[->, thick] (zU.north) to[out=90, in=180] 
            node[txt, pos=0.3] {$\mathcal{F}$} 
            node[txt, pos=0.7] {$\mathrm{\phi}$} (K.west);
            
        \draw[->, thick] (zU.south) to[out=-90, in=180] 
            node[txt, pos=0.3] {$\mathcal{F}$} 
            node[txt, pos=0.7] {$\mathrm{\phi}$} (V.west);

        \draw[->, thick] (K.east) to[out=0, in=90] (Gat1.north);
        \draw[->, thick] (K.east) to[out=0, in=90] (Gat2.north);
        \draw[->, thick] (K.east) to[out=0, in=90] (Gat3.north);

        \draw[->, thick] (V.east) to[out=0, in=-90] (Gat1.south);
        \draw[->, thick] (V.east) to[out=0, in=-90] (Gat2.south);
        \draw[->, thick] (V.east) to[out=0, in=-90] (Gat3.south);

        \end{tikzpicture}%
    }
    \caption{Schematic representation of the general gating-based architecture investigated in this work.}
    \label{fig:attention_arch}
\end{figure*}
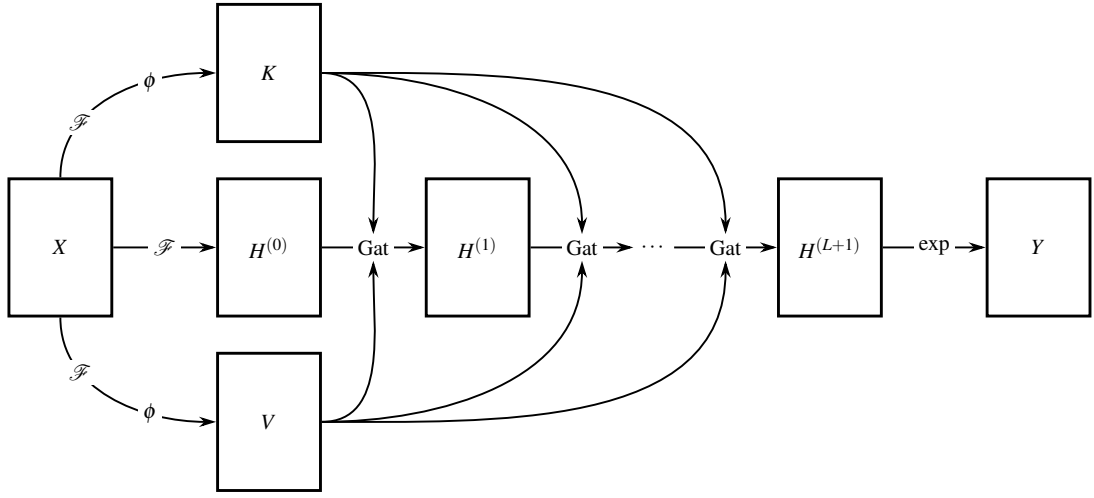

From $H^{(0)}$, the global encoder matrices $K,V \in \mathbb{R}^{n \times m}$ are computed using an activation function $\phi$ with the adaptive parameter $a^{(1)}$. In each subsequent layer $l$, the previous hidden state $H^{(l-1)}$ is used to compute a pre-activation $Z^{(l)} \in \mathbb{R}^{n \times m}$ via an affine transformation. An important component of this architectural structure lies in how the new hidden state $H^{(l)}$ is computed using a gating mechanism $\mathrm{Gat}(\cdot)$. This mechanism aggregates the pre-activation $Z^{(l)}$ with the global $K$ and $V$ matrices while explicitly incorporating the layer-specific adaptive parameter $a^{(l+1)}$. After iterating this process across $L$ layers, a final affine transformation is applied to the last hidden state $H^{(L)}$, yielding $H^{(L+1)}$. The network output $Y$ is obtained by applying the exponential function to the negative of this result~\cite{AlAradi2018}. The use of the exponential function ensures that the network's output remains strictly positive, thereby enforcing a hard physical constraint on the solution suitable for the problem. A schematic representation of this gating-based architecture is given by Fig.~\ref{fig:attention_arch}.

In Ref.~\onlinecite{Wang-2021-ID6448}, a type of gating mechanism was introduced that draws inspiration from the Gated Recurrent Unit (GRU) commonly employed in recurrent neural networks~\cite{Cho-2014-ID6453}. In the present work, a learnable parameter $a$ is introduced that scales the argument to the activation function adaptively. The mechanism takes the form
\begin{equation}
    \label{eq:wang_type_att1}
    \mathrm{Gat}_{\mathcal{W}}(a,Z, K, V) =  (1-\phi(aZ)) \odot K + \phi(aZ) \odot V,
\end{equation}
where $\odot$ denotes the Hadamard product. In the original formulation, $\phi(\cdot)=\mathrm{tanh}(\cdot)$ is used. As demonstrated in Ref.~\onlinecite{Wang-2021-ID6448}, this type of mechanism improves the performance of PINN training for idealized benchmark problems such as the Helmholtz and Klein-Gordon equations. However, it has been found in the present study that such a mechanism is not suitable for the present problem, largely due to vanishing gradients.

To overcome this, the present work proposes a scaled multiplicative gating mechanism that replaces the linear interpolation of Eq.~\eqref{eq:wang_type_att1} by a non-linear modulation scheme defined as
\begin{equation}
\label{eq:jannet_attention}
    \mathrm{Gat}_{\mathcal{A}}(a,Z, K, V) = \phi\left(a\frac{Z \odot K}{\sqrt{m}}\right) \odot V.
\end{equation}
This mechanism is conceptually inspired by the scaled dot-product attention of Ref.~\onlinecite{Vaswani2017}. The scaling factor $1/\sqrt{m}$ serves as a heuristic normalization to regulate the magnitude of the activation argument, and the learnable scalar $a$ allows the model to adaptively adjust this scaling. This ensures the input to $\phi$ remains within its 
active regime, preventing numerical saturation and mitigating the risk of vanishing gradients to facilitate stable training. Furthermore, the pure multiplicative form of the mechanism is crucial for the representation learning required in this context. An ablation study and validation of the neural network architecture with this gating mechanism are provided in Appendix~\ref{appendix:ablation_study} and Appendix~\ref{appendix:validation}.

\section{Experimental setup}

\begin{figure*}[htbp]
\centering
\includegraphics[width=0.8\textwidth]{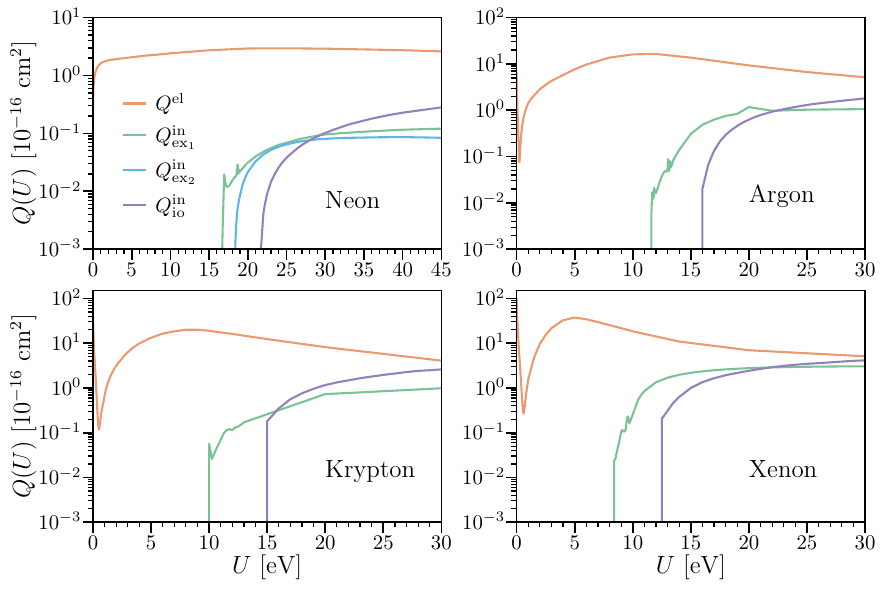}
\caption{\label{fig:cross_data}Electron collision cross-section data for neon, argon, krypton, and xenon.}
\end{figure*}

This work considers solving Eq.~\eqref{eq:final_equation} under the settings described in Sec.~\ref{sec:problem_setup} for four noble gases: neon, argon, krypton, and xenon. Figure~\ref{fig:cross_data} presents the electron collision cross-section data obtained from Ref.~\onlinecite{Hayashi-1992-ID751} for neon; Refs.~\onlinecite{Hayashi2003, Rapp-1965-ID5358} for argon; Refs.~\onlinecite{Fon-1981-ID6454, Mitroy-1990-ID6455, Specht-1980-ID6456, deHeer-1979-ID6457, Wetzel-1987-ID6458} for krypton; and Refs.~\onlinecite{Lam-1982-ID6459, Hayashi_1983, Puech-1991-ID6460, Wetzel-1987-ID6458} for xenon. Within the governing equations, the elastic momentum transfer cross section $Q^\mathrm{el}(U)$ and the inelastic $Q_k^\mathrm{in}(U)$ cross sections function as variable coefficients. Recall that the subscript $k$ denotes a specific inelastic process belonging to the set $\mathcal{P}$. For neon, $\mathcal{P}=\{\mathrm{ex}_1, \mathrm{ex}_2, \mathrm{io}\}$, representing two distinct excitation processes and one ionization process. For the other gases, the set is $\mathcal{P}=\{\mathrm{ex}_1, \mathrm{io}\}$. Notably, the sharp features inherent in these cross sections, such as the deep Ramsauer-Townsend minima in argon, krypton, and xenon, significantly complicate the training of PINNs.

The standard physical parameters generally shared across all calculations are listed in Table \ref{table:parameters}. A uniform electric field is applied in all cases. Furthermore, the maximum kinetic energy $U_\mathrm{max}$ is set to 45~eV, 25~eV, 20~eV, and 16~eV, for neon, argon, krypton, and xenon, respectively. These values were chosen to be consistent with the well-established finite difference numerical solver introduced in Ref.~\onlinecite{Loffhagen-2005-ID3906}, which is employed to produce all reference solutions used in this work.

\begin{table}[htbp]
\begin{threeparttable}
\caption{\label{table:parameters} Standard physical parameters generally used in the calculations.}
\centering
\begin{ruledtabular}
\begin{tabular}{lll}
 Physical parameter  & Description & Value \\ 
\hline
 $N$ & Gas number density\textsuperscript{\S} & 3.54$\times10^{16}$~cm$^{-3}$\\  
 $j$ & Electron current density & 2~mA/cm$^2$ \\
 $U_\mathrm{center}$ & Center energy & 5~eV \\
 $U_\mathrm{width}$ & Energy width & 2~eV \\
 $z_\mathrm{max}$ & Reactor length & 10~cm \\
 $\gamma$ & Reflection factor & 0.7\\
 $E$ & Electric field & $-5$~V/cm \\
 $T$ & Gas temperature & 273~K \\
\end{tabular}
     \begin{tablenotes}
     \small
       \item [\textsuperscript{\S}]The gas number density value corresponds to a gas pressure of 133.32~Pa at the given gas temperature.
     \end{tablenotes}
\end{ruledtabular}
\end{threeparttable}
\end{table}

The complexity of the problem places a critical importance on the choice of the neural network architecture used to approximate the solution $f_\theta(z,U)$. In this work, four feed-forward network architectures are investigated as given in Table \ref{table:architectures}. For every architecture, the trainable weights are initialized following the Glorot scheme~\cite{Glorot2010} and the adaptive parameters $a$ are initialized to one.

\begin{table}[htbp]
\caption{\label{table:architectures} Considered neural network architectures.}
\centering
\begin{ruledtabular}
\begin{tabular}{lp{3cm}p{3cm}}
 Architecture & Description & Notable component \\ 
\hline
 FCNNet & Standard fully connected architecture in Ref.~\onlinecite{Raissi-2019-ID6187} & --- \\
WangNet & Original gating-based architecture in Ref.~\onlinecite{Wang-2021-ID6448} & Gating mechanism in Eq.~\eqref{eq:wang_type_att1}\cite{Wang-2021-ID6448}, without adaptive activation functions (AAFs) \\  
 ModNet & Improved WangNet architecture & Fourier-feature layer~\cite{Tancik2020}, gating mechanism in Eq.~\eqref{eq:wang_type_att1} \cite{Wang-2021-ID6448}, AAFs~\cite{Jagtap2020} \\
 ANNet &  Another neural network architecture (present work) & Fourier-feature layer~\cite{Tancik2020}, gating mechanism in Eq.~\eqref{eq:jannet_attention}, AAFs~\cite{Jagtap2020} \\
\end{tabular}
\end{ruledtabular}
\end{table}

On top of architectural design, solving Eq.~\eqref{eq:final_equation} using the PINN framework requires a suitable formulation of the loss function. Here, one seeks to approximate the physical solution $f_0(z,U)$ by use of a neural network. To ensure numerical stability, the problem is formulated in a non-dimensional domain. To this end, a neural network $u_\theta(\tilde z, \tilde U)$ is introduced that approximates the dimensionless distribution function, where the inputs $\tilde z$ and $\tilde U$ are scaled to be of order unity by the scaling quantities $z_s$ and $U_s$, respectively (specified in Table~\ref{table:hyperparameters}). The relationship between the physical approximation $f_\theta(z,U)$ and the network output $u_\theta(\tilde z, \tilde U)$ is given by $f_\theta(z,U)=f_\mathrm{s}\cdot u_\theta(\tilde z, \tilde U)$, where $f_\mathrm{s}$ is a characteristic scaling factor to make the output of the neural network physical. The physical coordinates are related to the network inputs via the linear transformations $\tilde z = z/z_\mathrm{s}$ and $\tilde U = U/U_\mathrm{s}$.

The set of variable coefficients $\mathcal{Q}$ includes the non-dimensionalized energy-dependent cross sections for the elastic collision process $\tilde Q^\mathrm{el}(\tilde U)$ and a set of non-dimensionalized cross sections for the inelastic collision processes \mbox{$\tilde Q^\mathrm{in}(\tilde U)=\{\tilde Q^\mathrm{in}_\mathrm{ex_1}(\tilde U),\tilde Q^\mathrm{in}_\mathrm{ex_2}(\tilde U), \tilde Q^\mathrm{in}_\mathrm{io}(\tilde U)\}$}.

Substituting these transformations into Eq.~$\eqref{eq:final_equation}$ and rearranging it into the residual form yields the non-dimensionalized residual of the primary PDE $\mathcal{R}_\theta(\tilde z,\tilde U)$ taking the form
\begin{widetext}
\begin{equation}
\begin{aligned}
\label{eq:physics_residual}
    \mathcal{R}_\theta(\tilde z, \tilde U) =&\frac{\tilde U}{3} \left[-\frac{\tilde Q_\Sigma(\tilde U)}{\tilde N} \frac{\partial^2 u_\theta}{\partial \tilde z^2} + \frac{\tilde Q_\Sigma(\tilde U)}{\tilde N} \left( \frac{\mathrm{d} \widetilde{(e_0E(z))}}{\mathrm{d} \tilde z} \frac{\partial u_\theta}{\partial \tilde U} + \widetilde{e_0E(z)} \frac{\partial^2 u_\theta}{\partial \tilde z \partial \tilde U}\right) \right] - \frac{1}{3} \widetilde{e_0 E(z)} \left[-\frac{\tilde Q_\Sigma(\tilde U)}{\tilde N}\frac{\partial u_\theta}{\partial \tilde z} \right. \\
    &+ \left. \tilde Q_\Sigma(\tilde U)\frac{\widetilde{e_0 E(z)}}{\tilde N}\frac{\partial u_\theta}{\partial \tilde U}
    - \frac{\tilde U}{\tilde N} \frac{\mathrm{d} \tilde Q_\Sigma(\tilde U)}{\mathrm{d} \tilde U} \frac{\partial u_\theta}{\partial \tilde z} -  \frac{\tilde U}{\tilde N} \tilde Q_\Sigma(\tilde U)  \frac{\partial^2 u_\theta}{\partial \tilde z \partial \tilde U} + \frac{\tilde U}{\tilde N} \widetilde{e_0 E(z)}\frac{\mathrm{d} \tilde Q_\Sigma(\tilde U)}{\mathrm{d} \tilde U} \frac{\partial u_\theta}{\partial \tilde U} +  \frac{\tilde U}{\tilde N} \widetilde{e_0 E(z)} \tilde Q_\Sigma(\tilde U) \frac{\partial^2 u_\theta}{ \partial \tilde U^2}\right] \\
    &- 2\frac{m_\mathrm{e}}{M} \tilde N\left[2 \tilde U \tilde Q^\mathrm{el}(\tilde U)u_\theta + \tilde U^2 \frac{\mathrm{d} \tilde Q^\mathrm{el}(\tilde U)}{\mathrm{d} \tilde U} u_\theta + \tilde U^2 \tilde Q^\mathrm{el}(\tilde U) \frac{\partial u_\theta}{\partial \tilde U} + \widetilde{k_\mathrm{B} T} \left(2 \tilde U \tilde Q^\mathrm{el}(\tilde U)\frac{\partial u_\theta}{\partial \tilde U} + \tilde U^2 \frac{\mathrm{d} \tilde Q^\mathrm{el}(\tilde U)}{\mathrm{d} \tilde U}\frac{\partial u_\theta}{\partial \tilde U} \right. \right. \\
    &+ \left. \left. \tilde U^2 \tilde Q^\mathrm{el}(\tilde U) \frac{\partial^2 u_\theta}{\partial \tilde U^2}\right)\right] + \sum_{k \in \mathcal{P}} \tilde U \tilde N \tilde Q_k^\mathrm{in}(\tilde U) u_\theta - \sum_{k \in \mathcal{P}} (\tilde U + \tilde U_k^\mathrm{in} ) \tilde N \tilde Q_k^\mathrm{in}(\tilde U + \tilde U_k^\mathrm{in}) u_\theta(\tilde z, \tilde U + \tilde U_k^\mathrm{in}). 
\end{aligned}
\end{equation}
\end{widetext}
Here, tildes indicate dimensionless variables and constants. For brevity, the specific transformations are omitted. Each physical parameter is non-dimensionalized by applying the characteristic length $z_\mathrm{s}$ and energy $U_\mathrm{s}$ corresponding to its physical dimensions, yielding, for example, $\widetilde{e_0 E(z)} = e_0 E(z) (z_\mathrm{s} / U_\mathrm{s})$ and $\widetilde{k_\mathrm{B} T} = k_\mathrm{B} T / U_\mathrm{s}$ for the electric field force and thermal energy, respectively.

To formulate the optimization objective, an $L_1$-norm is applied to construct the PDE residual loss term as
\begin{equation}
\mathcal{L}_\mathrm{f}(\theta) = \frac{1}{n_\mathrm{f}} \sum_{i=1}^{n_\mathrm{f}} \| \mathcal{R}_\theta (\tilde z_\mathrm{f}^{(i)}, \tilde U_\mathrm{f}^{(i)}) \|_1. 
\end{equation}
Latin Hypercube Sampling is used to sample a new set of PDE residual collocation points $\{(\tilde z_\mathrm{f}^{(i)}, \tilde U_\mathrm{f}^{(i)})\}_{i=1}^{n_\mathrm{f}}$ at each training iteration.

Following the general form of Eq.~\eqref{eq:general_loss}, the total loss function is a composite loss of a weighted sum of the PDE residual loss and the boundary condition losses. For the present numerical implementation using PINNs, it has been found empirically that excluding the explicit boundary condition at the upper energy limit $(z,U\geq U_\mathrm{max})$ improves numerical stability. Because inelastic collisions physically dictate a rapid, exponential decay of the high-energy tail, the condition that the electron population becomes negligible at high energies is intrinsically captured by the PDE residual, provided $U_\mathrm{max}$ is chosen sufficiently large for the given gas and electric field. Consequently, the total loss can be written as
\begin{equation}
\begin{aligned}
\label{eq:total_loss}
\mathcal{L}(\theta) = &w_\mathrm{f}\mathcal{L}_\mathrm{f}(\theta) + w_\mathrm{a}\mathcal{L}_\mathrm{a}(\theta) \\&+ w_\mathrm{b}\mathcal{L}_\mathrm{b}(\theta) + w_\mathrm{c}\mathcal{L}_\mathrm{c}(\theta) + w_\mathrm{d}A(a),
\end{aligned}
\end{equation}
where the boundary loss terms are given by
\begin{alignat}{2}
    \mathcal{L}_\mathrm{a}(\theta) &= \frac{1}{n_\mathrm{a}} \sum_{i=1}^{n_\mathrm{a}} \|u_{\theta1}( \tilde z_\mathrm{max}, \tilde U_\mathrm{a}^{(i)}) - \gamma u_\theta(\tilde z_\mathrm{max}, \tilde U_\mathrm{a}^{(i)})\|_1, &\,& (\mathrm{anode})\\
    \mathcal{L}_\mathrm{b}(\theta) &= \frac{1}{n_\mathrm{b}} \sum_{i=1}^{n_\mathrm{b}} \|u_{\theta1}(\tilde z_\mathrm{b}^{(i)},0)\|_1, &\,& (\mathrm{bottom})\\
    \mathcal{L}_\mathrm{c}(\theta) &= \frac{1}{n_\mathrm{c}} \sum_{i=1}^{n_\mathrm{c}} \|u_{\theta1}(0,\tilde U_\mathrm{c}^{(i)}) - \tilde g(\tilde U_\mathrm{c}^{(i)})\|_1, &\,& (\mathrm{cathode})     
\end{alignat}
and \(A(a)=(1/L \sum_{l=1}^L \mathrm{exp}[a^{(l)}])^{-1}\) is the slope recovery term for the adaptive activation functions \cite{Jagtap2020}. In these loss terms, $u_{\theta1}(\tilde{z}, \tilde{U})$ represents the dimensionless anisotropic component of the EVDF. It is computed from the network output $u_\theta$ via automatic differentiation according to the dimensionless form of Eq.~\eqref{eq:f_1}.

It has been found that excluding $\mathcal{L}_\mathrm{a}(\theta)$ in the early stages of training is crucial for convergence. Consequently. this term is introduced only after a sufficient \textit{warm-up} period, e.g.\ around iteration $100 \times 10^3$. The total loss $\mathcal{L}(\theta)$ is minimized via the Adam optimizer~\cite{Kingma2015} with an exponential learning rate schedule that decays from $10^{-3}$ initially to $10^{-4}$ at iteration $10 \times 10^{6}$. Table \ref{table:hyperparameters} summarizes the hyperparameters used for all training cases. All PINN architectures employ the hyperbolic tangent ($\tanh$) activation function. The computational framework is implemented in PyTorch v2.6.0~\cite{Paszke2019} on an Nvidia A100 80GB graphics processing unit. Training is performed using a 32-bit floating-point precision (FP32); although 64-bit precision (FP64) yields a slight increase in accuracy, it is not required for convergence in the given cases, and the marginal gain does not justify the additional computational cost and doubled memory requirements.

\begin{table}[ht]
\centering
\caption{\label{table:hyperparameters} Hyperparameters used in all PINN training cases.}
\begin{ruledtabular}
\begin{tabular}{lp{4.5cm}p{1.5cm}}
 Hyperparameter  & Description & Value \\ 
\hline
 $m$ & Number of neurons in each hidden layer & 200 \\  
 $L$ & Number of hidden layers & 6  \\
 $f_\mathrm{s}$ & Output scaling quantity & $1.7\times 10^{9}$ eV$^{-3/2}$/cm$^3$\\
 $z_s$ & $z$ scaling quantity & 10 cm \\
 $U_s$ & $U$ scaling quantity & $U_\mathrm{max}$ \\
 $\sigma$ & Standard deviation for Fourier-feature layer & 10 \\
  $n_\mathrm{f}$ & Number of PDE residual collocation points & 8000 \\ 
 $n_\mathrm{a}$, $n_\mathrm{b}$, $n_\mathrm{c}$ & Number of boundary collocation points & 1000 \\
 $w_\mathrm{f}$ & PDE residual loss term weight & 0.7 \\
 $w_\mathrm{a}$ & Anode BC loss term weight & 0.01 \\
 $w_\mathrm{b}$, $w_\mathrm{c}$ & Bottom and cathode BC loss term weights & 0.29 \\
  $w_\mathrm{d}$ & Slope recovery term weight & 1 \\
\end{tabular}
\end{ruledtabular}
\end{table}

\section{Results and discussion}
This section presents and discusses the results obtained from applying the proposed ANNet architecture to the spatially one-dimensional electron Boltzmann equation. First, the convergence behavior of the four considered network architectures is analyzed, highlighting the role of gradient flow in achieving physically consistent solutions. The accuracy of the converged solutions is then evaluated for neon, argon, krypton, and xenon at standard conditions, both at the level of the isotropic distribution function and the derived macroscopic electron properties. Finally, the robustness of the architecture is assessed by varying the electric field strength using argon as an example, examining the limits of the current configuration and setup.

\subsection{Convergence analysis of PINN architectures}
\begin{figure}[htbp] 
    \centering
    
    \begin{subfigure}[t]{0.95\linewidth}
        \setcounter{subfigure}{0} 
        \centering
        \caption{$L_1$ total Loss, $\mathcal{L}(\theta)$}
        \includegraphics[width=\linewidth]{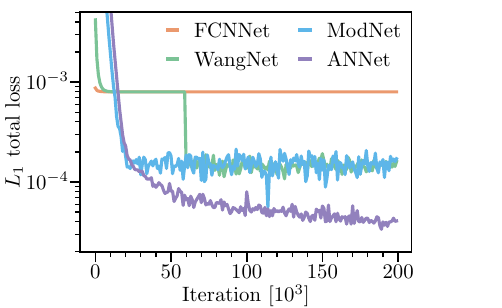}
        \label{fig:total_loss}
    \end{subfigure}\\
    \vspace{-0.3cm}
    \begin{subfigure}[t]{0.95\linewidth}
        \setcounter{subfigure}{1} 
        \centering
        \caption{ANNet, $f_\theta(z,U)$}
        \includegraphics[width=\linewidth]{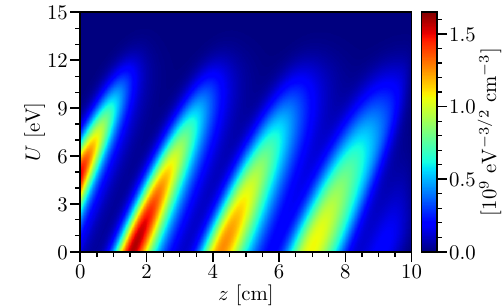}
        \label{fig:annet_sol}
    \end{subfigure}\\
    \vspace{-0.3cm}
    \begin{subfigure}[t]{0.95\linewidth}
        \setcounter{subfigure}{2} 
        \centering
        \caption{Reference, $f_0(z,U)$}
        \includegraphics[width=\linewidth]{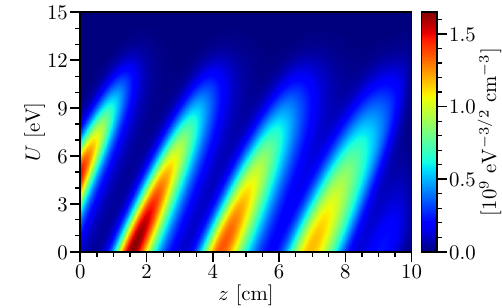}
        \label{fig:reference_sol}
    \end{subfigure}\\

    \caption{\label{fig:convergence_behavior}Convergence behavior of the considered network architectures for argon at standard conditions and the reference solution.}
\end{figure}

As noted in Sec.~\ref{sec:network_architecture}, training PINNs on complex equations presents a significant challenge, often leading to non-physical or trivial solutions. Solving Eq.~\eqref{eq:final_equation} for $f_0(z,U)$ strongly exemplifies these difficulties. Figure~\ref{fig:convergence_behavior} presents the solutions $f_\theta(z,U)$ obtained by the four considered architectures given in Table~\ref{table:architectures} for argon at standard conditions according to Table~\ref{table:parameters} and the hyperparameters provided by Table~\ref{table:hyperparameters}. The initial set of results refers to the training after 200$\times 10^3$ iterations and are compared against the reference solution. Ideally, the solution exhibits spatial striations, a characteristic feature of kinetic electron transport at this physical condition.

As shown in Fig.~\ref{fig:total_loss}, the loss evolution reveals distinct training dynamics for each architecture. The standard FCNNet gets trapped in a poor local minimum immediately in the training phase. A similar initial stagnation can be observed for WangNet. While it eventually escapes the saddle point, it settles into another poor local minimum. ModNet displays a rapid reduction in loss but subsequently gets stuck in an inadequate local minimum with a very fluctuating behavior. These suboptimal loss plateaus correspond to near-trivial solutions in which the distribution is effectively zero throughout the domain, failing to capture any of the physically expected features. In contrast, the proposed ANNet architecture demonstrates a stable decrease in loss as it converges toward a physically meaningful solution (Fig.~\ref{fig:annet_sol}), capturing the expected striations (Fig.~\ref{fig:reference_sol}).
\newlength{\subfiggap}
\setlength{\subfiggap}{-0.8cm}
\newcommand{\figscaleb}{0.45} 
\newcommand{\midgap}{7.3cm}  
\newlength{\totalwidth}
\setlength{\totalwidth}{\dimexpr \figscaleb\linewidth + \figscaleb\linewidth + \midgap \relax}

\begin{figure}[htbp] 
    \centering
    
    \begin{subfigure}[t]{0.9\linewidth}
        \setcounter{subfigure}{0} 
        \centering
        \includegraphics[width=\linewidth]{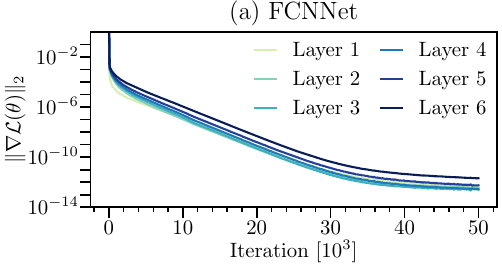}
        \label{fig:grad_fcnnet}
    \end{subfigure}\\
    \vspace{-0.2cm}
    \begin{subfigure}[t]{0.9\linewidth}
        \setcounter{subfigure}{1} 
        \centering
        \includegraphics[width=\linewidth]{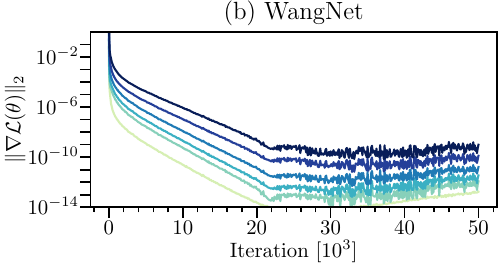}
        \label{fig:grad_wangnet}
    \end{subfigure}\\
    \vspace{-0.2cm}
    \begin{subfigure}[t]{0.9\linewidth}
        \setcounter{subfigure}{2} 
        \centering
        \includegraphics[width=\linewidth]{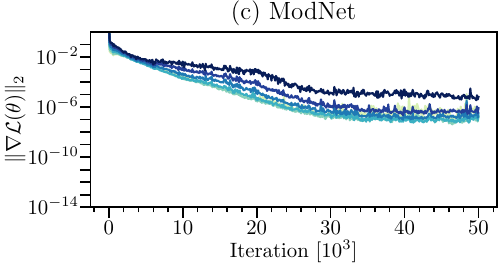}
        \label{fig:grad_modnet}
    \end{subfigure}\\
    \vspace{-0.2cm}
    \begin{subfigure}[t]{0.9\linewidth}
        \setcounter{subfigure}{3} 
        \centering
        \includegraphics[width=\linewidth]{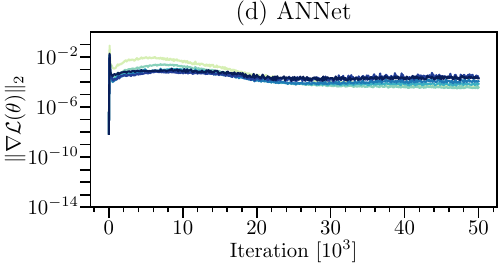}
        \label{fig:grad_annet}
    \end{subfigure}
    
    \setcounter{subfigure}{4}
    
    \caption{\label{fig:gradflows}$L_2$-norm of the back-propagated gradients with respect to the parameters of each hidden layer $\|\nabla \mathcal{L}(\theta) \|_2$ for argon at standard conditions.}
\end{figure}

Further analysis of the gradient flow emphasizes the critical role of initial convergence behavior (e.g.\ up to iteration $50\times10^3$) in obtaining a physically consistent solution. Here, an optimal convergence behavior can be characterized by two key indicators: (i) the magnitude of back-propagated gradients remains non-vanishing throughout training, and (ii) it is consistent across all hidden layers, ensuring that each hidden layer actively learns at the same rate. This favorable convergence behavior can be observed in ANNet as shown in Fig.~\ref{fig:gradflows}. Here, the $L_2$-norm of the back-propagated gradient of each hidden layer is shown as a function of the number of iterations.

In contrast, the other architectures suffer from pathological gradient flows. FCNNet and WangNet experience rapid gradient collapse, while ModNet shows a vanishing gradient flow throughout the hidden layers, i.e. the $L_2$-norm magnitude in layer 6 (closest to the output) becomes smaller as it flows to layer 1 (closest to the input). Although WangNet demonstrates excellent convergence and training stability on various standard benchmark PDEs~\cite{Wang-2021-ID6448}, the present results indicate that its gating mechanism formulation is ill-suited for the problem presented in this work. Notably, while the inclusion of Fourier features (ModNet, ANNet) generally improves overall gradient stability, the structural robustness in ANNet's gating mechanism is required to maintain a healthy and uniform gradient flow. As detailed in Appendix~\ref{appendix:varying_sigma}, ANNet demonstrates robustness provided the standard deviation for Fourier-feature layer $\sigma$ is appropriately tuned. ModNet does not show this robustness and consistently fails to yield a physical solution irrespective of $\sigma$.

\setlength{\subfiggap}{-0.8cm}
\setlength{\totalwidth}{\dimexpr \figscaleb\linewidth + \figscaleb\linewidth + \midgap \relax}

\subsection{Convergence rate and solution accuracy}
To quantify the accuracy of the learned solutions and derived quantities, the $L_2$ relative error $\varepsilon_{L_2} = \|\hat y-y\|_2 / \|y \|_2$ is defined for a learned quantity $\hat y$ with respect to a reference $y$ obtained from the finite difference numerical solver~\cite{Loffhagen-2005-ID3906}.

ANNet yields a physically reasonable solution for argon at standard conditions after $200 \times 10^3$ iterations achieving $\varepsilon_{L_2} \approx 6.08\times 10^{-2}$. The solution accuracy can be improved by extended training. A slow but stable decrease in error is observed reaching $\varepsilon_{L_2} \approx 2.06\times 10^{-2}$ after $10\times 10^6$ iterations. It is important to note that ANNet is trained separately from scratch for each individual gas case. The convergence behavior varies across gases and depends on the quantity of interest. For neon, macroscopic properties saturate well before the distribution itself, while in other cases certain macroscopic properties continue to improve even after the distribution has effectively converged. For reference, training for $200 \times 10^3$ iterations requires several hours of training, whereas $10\times 10^6$ iterations require approximately one week on a single NVIDIA A100 GPU.

\begin{figure*}[ht] 
    \centering
    
    \begin{subfigure}[t]{0.99\linewidth}
        \setcounter{subfigure}{0} 
        \centering
        \caption{Learned isotropic distribution by ANNet, $f_\theta$ [$10^{9}$ eV$^{-3/2}$ cm$^{-3}$]}
        \includegraphics[width=\linewidth]{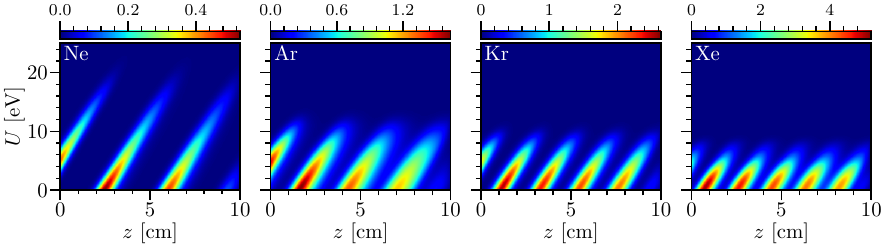}
        \label{fig:iso_dist}
    \end{subfigure}\\
    \vspace{-0.4cm}
    \begin{subfigure}[t]{0.99\linewidth}
        \setcounter{subfigure}{1} 
        \centering
        \caption{Max-normalized error, $|f_\theta - f_0| / \mathrm{max}(f_0)$}
        \includegraphics[width=\linewidth]{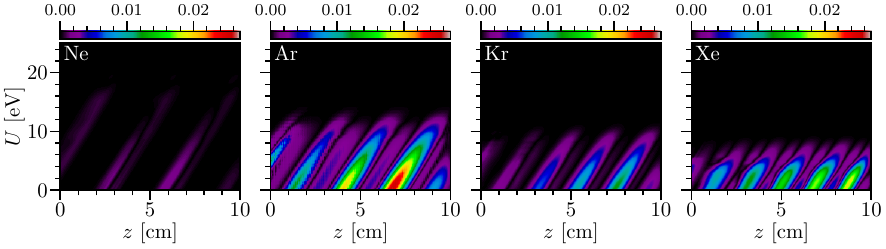}
        \label{fig:maxnorm_error}
    \end{subfigure}\\
\caption{\label{fig:converged_solutions}\textbf{Top row}: Isotropic distributions $f_\theta(z,U)$ learned by the ANNet architecture for neon (Ne), argon (Ar), krypton (Kr), and xenon (Xe) after $10 \times 10^6$ iterations at standard conditions. \textbf{Bottom row}: Max-normalized error maps for ANNet in comparison to the reference solutions.}
\end{figure*}

The top row of Fig.~\ref{fig:converged_solutions} shows the converged isotropic distribution $f_\theta(z,U)$ obtained by ANNet after $10\times 10^6$ iterations for neon, argon, krypton, and xenon, respectively, at standard conditions given in Table~\ref{table:parameters}. Using the fixed hyperparameters listed in
Table~\ref{table:hyperparameters}, the architectural setup successfully generalizes across all considered gases, achieving final $\varepsilon_{L_2}$ values of $0.17 \times 10^{-2}$ (neon), $2.06 \times10^{-2}$ (argon), $0.92 \times10^{-2}$ (krypton), and $1.69 \times10^{-2}$ (xenon).

As further shown in the bottom row of Fig.~\ref{fig:converged_solutions}, the error does not follow a strictly monotonic trend with respect to the striation frequency, which is inversely proportional to the period length. Neural networks are known to exhibit spectral bias~\cite{Rahaman2019}, which is a tendency to favor the learning of low-frequency features over high-frequency ones. The finding here deviates from this expectation. Notably, argon displays higher error in its maxima than xenon, despite having a less-striated solution structure (lower frequency). This suggests that the achievable accuracy is not determined solely by spatial frequency, but rather by a complex interplay between the collisional cross-section data, the inherent spectral bias, and the choice of hyperparameters. Since all solutions were obtained using a fixed set of hyperparameters to demonstrate architectural robustness, case-specific tuning might further reduce errors. The maximum of max-normalized error and $\varepsilon_{L_2}$ always remain below 3\% for all cases, confirming the architecture's stability and generalization capability.

\setlength{\subfiggap}{-0.8cm}
\setlength{\totalwidth}{\dimexpr \figscaleb\linewidth + \figscaleb\linewidth + \midgap \relax}

To subject the learned solutions to closer scrutiny, Fig.~\ref{fig:slice_plots} shows the isotropic distribution normalized by the electron number density, $f_0(z,U) / n_\mathrm{e}(z)$. This normalized distribution is plotted as a function of the kinetic energy $U$, at positions $z$ = 1, 5, and 9 cm. A logarithmic scale is used here; on this scale, a Maxwellian distribution corresponds to a linearly decreasing function for increasing $U$. The plots demonstrate that the learned solutions generally maintain excellent agreement over a substantial dynamic range extending down to $\sim10^{-7}$~eV$^{-3/2}$ for all gases. Minor exceptions occur for neon at $z=1$~cm, specifically around $U\approx 2$~eV, and at $z=5$~cm around $U\approx 22$~eV. These discrepancies arise because the neon solution exhibits steeper profiles along the kinetic energy axis $U$ than the other rare gases. The latter show high-frequency oscillations primarily along the spatial axis $z$.

\def\figscale{0.45} 
\def\midgap{7cm}
\setlength{\totalwidth}{\dimexpr \figscale\linewidth + \figscale\linewidth + \midgap \relax}
\begin{figure*}[ht]
    \centering
    \makebox[\totalwidth][l]{%
        \makebox[\totalwidth][r]{%
            \begin{subfigure}[t]{\figscale\linewidth}
                \setcounter{subfigure}{1} 
                \centering
                \includegraphics[width=\linewidth]{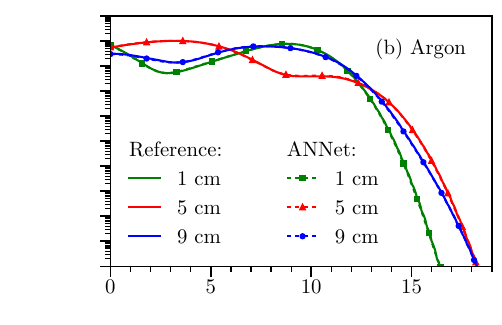}
                \label{fig:slice_ar}
            \end{subfigure}%
        }%
        
        \hspace{-\totalwidth}%
        
        \makebox[\totalwidth][l]{%
            \begin{subfigure}[t]{\figscale\linewidth}
                \setcounter{subfigure}{0} 
                \centering
                \includegraphics[width=\linewidth]{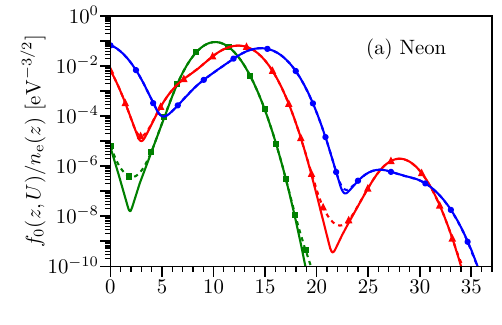}
                \label{fig:slice_ne}
            \end{subfigure}%
        }%
    }\\[-4ex]
    
    \makebox[\totalwidth][l]{%
        \makebox[\totalwidth][r]{%
            \begin{subfigure}[t]{\figscale\linewidth}
                \setcounter{subfigure}{3} 
                \centering
                \includegraphics[width=\linewidth]{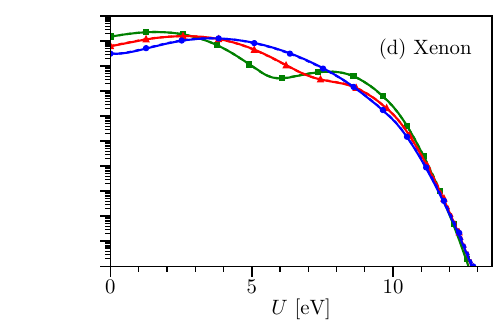}
                \label{fig:slice_xe}
            \end{subfigure}%
        }%
        
        \hspace{-\totalwidth}%
        
        \makebox[\totalwidth][l]{%
            \begin{subfigure}[t]{\figscale\linewidth}
                \setcounter{subfigure}{2} 
                \centering
                \includegraphics[width=\linewidth]{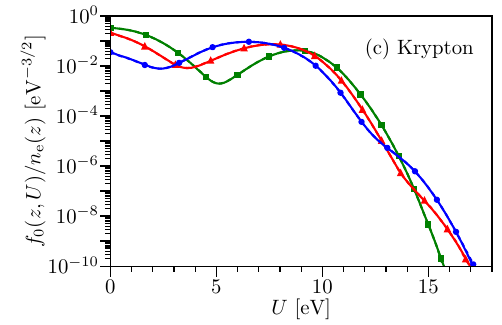}
                \label{fig:slice_kr}
            \end{subfigure}%
        }%
    }\\[-3ex]
    \setcounter{subfigure}{4}
    
    \caption{\label{fig:slice_plots}Logarithmic slice plots of the isotropic distribution $f_0 (z,U)$ normalized by the local electron densities $n_\mathrm{e}(z)$ obtained from Eq.~\eqref{eq:electron_density}. The plots compare the reference solutions (solid lines) with the ANNet solutions (dashed lines with symbols) at $z=$ 1, 5, and 9 cm across the four considered gases at standard conditions.}
\end{figure*}

\subsection{Accuracy of macroscopic electron properties}

\begin{figure*}[htbp]
    \centering
    \def\figscale{0.4} 
    
    \begin{subfigure}[t]{\figscale\linewidth}
        \centering
        \caption{Neon}
        \includegraphics[width=\linewidth]{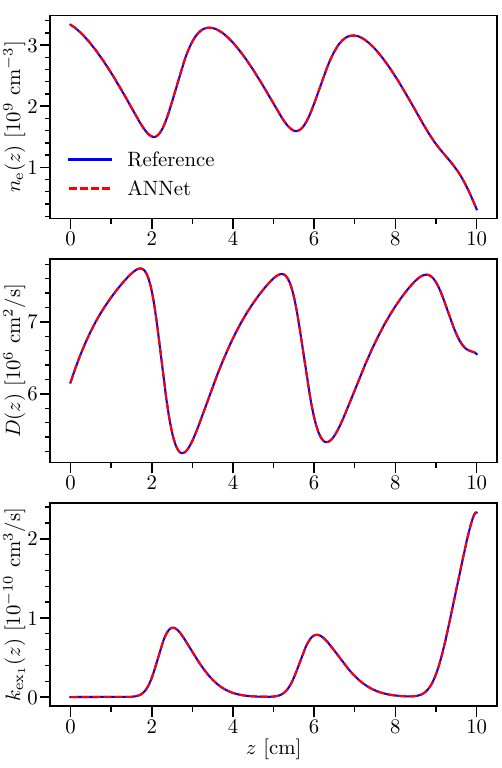}
        \label{fig:macro_neon}
    \end{subfigure}
    \hspace{-0.2cm} 
    \begin{subfigure}[t]{\figscale\linewidth}
        \centering
        \caption{Argon}
        \includegraphics[width=\linewidth]{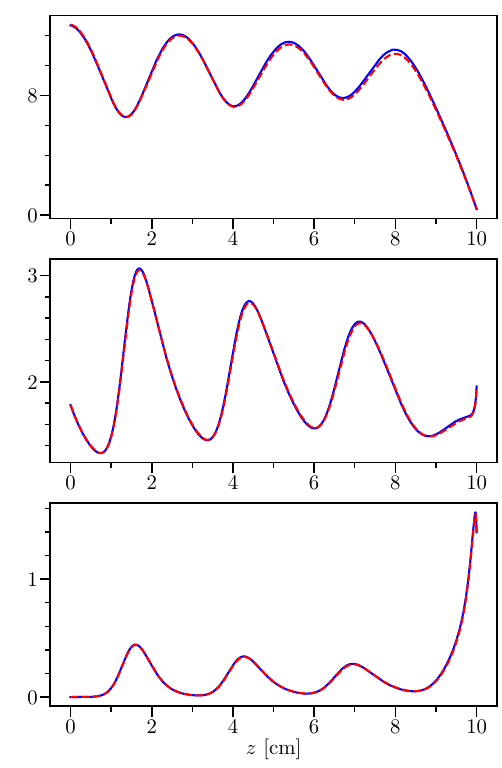}
        \label{fig:macro_argon}
    \end{subfigure}
    \caption{\label{fig:macro_quantities}Comparison of selected macroscopic electron properties: electron density $n_\mathrm{e}(z)$, diffusion coefficient $D(z)$, and excitation rate coefficient $k_{\mathrm{ex}_1}(z)$, for neon and argon at standard conditions.}
\end{figure*}

The accuracy of the present approach using ANNet relating to correct macroscopic physics is further investigated. Figure~\ref{fig:macro_quantities} presents the spatial evolution of selected macroscopic quantities for neon and argon. Specifically, the figure compares the electron density $n_\mathrm{e}(z)$, diffusion coefficient $D(z)$, and excitation rate coefficient $k_{\mathrm{ex}_1}(z)$. These properties are derived from the reference isotropic distribution $f_0(z,U)$ and the learned PINN distribution $f_\theta(z,U)$ using Eqs.~\eqref{eq:electron_density}, \eqref{eq:diffusion_coeff}, and \eqref{eq:rate_coeff}, respectively. For simplicity, the same notation is used for macroscopic properties regardless of the underlying solution method. The results demonstrate excellent agreement between the ANNet solutions and the reference solutions. Note that the results for krypton and xenon exhibit the same agreement quality. Importantly, ANNet accurately reproduces the characteristic spatial relaxation oscillations, i.e. the periodic rise and fall of electron properties caused by the cyclic gain of energy from the electric field and subsequent loss via inelastic collisions.

\begin{figure*}[htbp]
\centering
\includegraphics[width=0.8\textwidth]{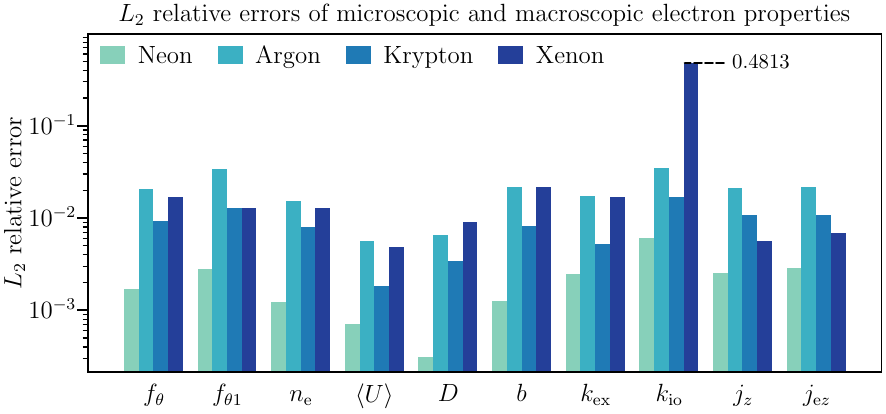}
\caption{\label{fig:barchart}$L_2$ relative errors $\varepsilon_{L_2}$ of the isotropic and anisotropic distribution (microscopic properties) as well as the derived macroscopic electron properties across all considered gases at standard conditions. For neon, the value is the average of $\varepsilon_{L_2}(k_\mathrm{ex_1})$ and $\varepsilon_{L_2}(k_\mathrm{ex_2})$, which are $0.13 \times 10^{-2}$ and $0.36 \times 10^{-2}$, respectively. For the other gases, $k_\mathrm{ex} := k_\mathrm{ex_1}$.}
\end{figure*}

Figure~\ref{fig:barchart} further details $\varepsilon_{L_2}$ values for all considered quantities. Here, the anisotropic distribution $f_{\theta1}(z,U)$ is evaluated using Eq.~\eqref{eq:f_1} by substituting $f_0(z,U)$ with $f_\theta(z,U)$, while the electron mean energy $\langle U \rangle$, particle flux $j_z(z)$, energy flux $j_{\mathrm{e}z}(z)$, mobility $b(z)$, and ionization rate coefficient $k_\mathrm{io}(z)$ are calculated following Eqs.~\eqref{eq:mean_energy}, \eqref{eq:particle_flux}, \eqref{eq:energy_flux}, \eqref{eq:mobility}, and \eqref{eq:rate_coeff}, respectively. It reveals that $k_\mathrm{io}(z)$ consistently has higher relative errors compared to other macroscopic properties, most notably in the xenon case.

The ionization rate is physically determined specifically by the high-energy tail of the distribution function, a region where the isotropic distribution has already decreased by several orders of magnitude relative to its maximum value. Resolving this higher energy tail presents a significant numerical challenge for the neural network approach, as the population density in this region almost vanishes compared to the low-energetic electrons. The cause of this difficulty is twofold: (i) numerically, it is linked to the spectral bias of neural networks; and (ii) physically, it is highly influenced by the strict energy thresholds governed by the collision cross-section data.

This issue is exacerbated for xenon by the extreme disparity in magnitude between $k_\mathrm{ex_1}(z)\sim 10^{-10}$ cm$^3$/s and $k_\mathrm{io}(z)\sim 10^{-17}$ cm$^3$/s. The network struggles to optimize the ionization contribution without introducing numerical artifacts. As noted previously, transitioning from FP32 to FP64 can partially mitigate these numerical artifacts in the stiff high-energy regime. However, the marginal increase in accuracy for properties like $k_\mathrm{io}(z)$ does not justify the doubled memory requirements and increased computational cost. Despite these quantitative discrepancies in the high-energy tail, ANNet generally succeeds in recovering the global spatial periodicity of the ionization profile, capturing the underlying physical structure even in this numerically challenging and stiff regime.

\subsection{Variation of electric field strength}
To assess the robustness of the ANNet architecture with respect to variations in physical parameters, Fig.~\ref{fig:3dplots} presents the normalized learned isotropic distribution function $f_\theta(z,U)/n_\mathrm{e}(z)$ on a logarithmic scale for argon at $E$ fields of $-8$, $-5$, and $-2$~V/cm. Other physical parameters and hyperparameters remain the same as found in Table~\ref{table:parameters} and Table~\ref{table:hyperparameters}, respectively. These solutions are obtained from independent cases trained up to $10 \times 10^6$ iterations. Additionally, the normalized anisotropic distribution $f_1(0,z)/n_\mathrm{e}(0)$ representing the electron influx at $z=0$ is depicted as a black line.

Physically, the electric field magnitude $|E|$ governs the energy gained by electrons between elastic and inelastic collisions. The latter mainly dictates the spatial frequency and amplitude of the relaxation oscillations. It is found that a larger field magnitude ($|E|=8$~V/cm) induces steeper energy gain, which results in a prominently striated structure characterized by high-frequency spatial features. In contrast, reducing the magnitude to \mbox{$|E|=2$~V/cm} extends the spatial period length, yielding a smoother distribution where the characteristic striations are significantly dampened due to elastic collision processes.

The $L_2$ relative errors ($\varepsilon_{L_2}$) are $2.23 \times 10^{-2}$, $ 2.06\times10^{-2}$, and $4.39\times10^{-2}$ for $E =-8$, $-5$, and $-2$ V/cm, respectively. It reveals a non-monotonic trend, with the lowest error observed at the intermediate field magnitude of \mbox{$|E|=5$~V/cm}. This variation in accuracy coincides with the changing physical characteristics of the domain across the evaluated cases. In the high-magnitude regime ($|E|=8$~V/cm), the solution exhibits high-frequency spatial topologies. Conversely, at the lowest field magnitude ($|E|=2$~V/cm), reflected electrons penetrate significantly deeper into the bulk ($\lambda \approx 5.7$~cm\footnote{The period length is given by $\lambda\approx U^\mathrm{in}_\mathrm{ex_1} / \left(e_0|E|\right)$, where $U^\mathrm{in}_\mathrm{ex_1}\approx11.55$~eV is the lowest excitation threshold energy for argon.}) compared to the $|E|=8$~V/cm case ($\lambda\approx1.4$~cm), thereby broadening the spatial influence of the anode boundary condition. The overall error remains stable and within the same order of magnitude across the different field strengths.

Notably, these distinct solutions were obtained using the identical hyperparameters presented in Table~\ref{table:hyperparameters}, demonstrating the architecture's robustness across varying solution topologies without case-specific retuning. Extending to larger electric field magnitudes (e.g.\ $|E|\geq10$~V/cm) exceeds the capacity of the present configuration, making the network prone to convergence failure. Addressing these highly oscillatory and stiff regimes necessitates explicit hyperparameter adjustments or advanced training strategies, establishing a practical boundary for the current setup and a direction for future work.

\def\figscale{0.32} 
\begin{figure*}[htbp] 
    \centering
    
    \begin{subfigure}[t]{\figscale\linewidth}
        \centering
        \includegraphics[width=\linewidth]{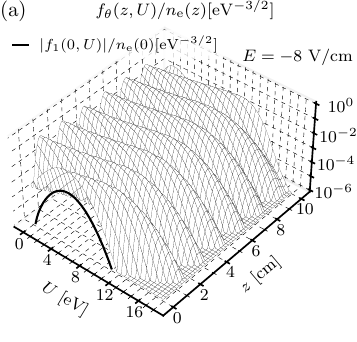}
        \label{fig:loss}
    \end{subfigure}%
    \hfill 
    \begin{subfigure}[t]{\figscale\linewidth}
        \centering
        \includegraphics[width=\linewidth]{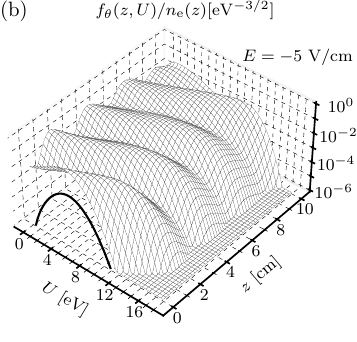}
        \label{fig:wangnet}
    \end{subfigure}%
    \hfill
    \begin{subfigure}[t]{\figscale\linewidth}
        \centering
        \includegraphics[width=\linewidth]{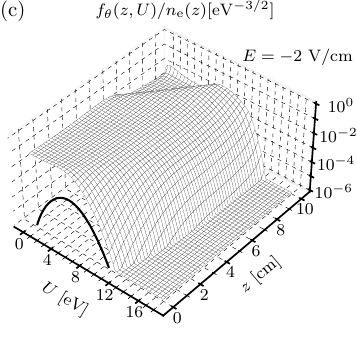}
        \label{fig:annet}
    \end{subfigure}
    \caption{\label{fig:3dplots} Surface plots of the normalized learned isotropic distribution $f_\theta(z,U)/n_\mathrm{e}(z)$ on a logarithmic scale for argon at different electric field values. The solutions are obtained using the same set of hyperparameters as in Table~\ref{table:hyperparameters}.}
\end{figure*}

\section{Conclusion and outlook}
The present work investigates the applicability of \textit{physics-informed neural networks} (PINNs) for solving the 1D spatially inhomogeneous electron Boltzmann equation subject to a uniform electric field for neon, argon, krypton, and xenon. The equation is treated within the two-term approximation framework of the Legendre polynomial expansion of the EVDF. The resulting PDE for the isotropic part of the EVDF is solved directly in kinetic energy space without the conventional transformation to total energy typically employed by direct numerical solution methods such as finite-difference schemes. This highlights the flexibility of the PINN framework regarding the choice equation formulations. Resolving the isotropic and anisotropic part of the EVDF is crucial for the accurate determination of nonlocal macroscopic electron properties relevant for low-temperature plasma simulations.

The neural network architecture ANNet with a new gating mechanism is introduced. It overcomes the problem resulting from other recent architectures, which lead to physically incorrect or trivial solutions. The architecture is characterized by a Fourier-feature input layer, adaptive activation functions, and a purely multiplicative gating mechanism with scaling. The results demonstrate that this formulation is crucial for maintaining healthy non-vanishing gradient flow throughout network layers, thereby facilitating convergence to physically correct solutions. The ANNet solutions are contrasted against reference solutions, where ANNet is shown to be capable of producing solutions with excellent agreements across various microscopic (isotropic and anisotropic distribution) and derived macroscopic quantities. Furthermore, the proposed architecture exhibits robustness that generalizes effectively across different gas types and a defined range of electric field strengths without requiring case-specific hyperparameter tuning.

The present PINN approach still faces the significant challenge related to computational efficiency, as achieving accurate solutions requires training up to $10 \times 10^6$ iterations. Furthermore, scaling the approach to stronger electric fields has proven difficult; addressing these complexities provides a clear opportunity for further algorithmic refinement.

Future work will investigate suitable optimization and training strategies tailored to expand the applicable range of field strengths and accelerate convergence for such complex PDEs. Overcoming these challenges will pave the way for incorporating greater physical complexity. Because the framework operates directly in kinetic energy space and bypasses the restrictive total energy transformation, it naturally accommodates scenarios that challenge conventional solvers, such as electric field reversals and multi-dimensional spatial domains.

\section*{Acknowledgment }
J.T. acknowledges support from Tokyo Electron Technology Solutions Ltd.

\section*{Data and code availability}
The data and code supporting this study are currently undergoing final curation and documentation for public release. They will be deposited on \url{https://zenodo.org/} and made openly available no later than six months after publication. Prior to final public release, access can be requested by contacting the corresponding author.

\appendix
\section*{Appendix}
\label{appendix}

\subsection{Classification of the two-term approximated spatially inhomogeneous electron Boltzmann equation}
\label{appendix:degenerate_elliptic}
This section presents the classification of the equation governing the isotropic distribution function. The equation is a second-order partial differential equation represented in Eq.~\eqref{eq:final_equation}. Let us focus on the principal part, which is determined by the coefficients of the highest-order derivatives.

By collecting the terms associated with $\partial^2 f_0/\partial z^2$, $\partial^2 f_0/\partial z \partial U$, and $\partial^2 f_0/\partial U^2$, we can represent the principal part using the coefficient matrix
\begin{equation}
    \begin{aligned}
    \mathbf{M} &= \begin{pmatrix} A & B \\ B & C \end{pmatrix}\\
    &=
    \begin{pmatrix}
        -\dfrac{U Q_\Sigma(U)}{3N} & \dfrac{e_0 E(z) U Q_\Sigma(U)}{3N} \\[1em]
        \dfrac{e_0 E(z) U Q_\Sigma(U)}{3N} & -\dfrac{(e_0 E(z))^2 U Q_\Sigma(U)}{3N} - \alpha(U)
    \end{pmatrix},
    \end{aligned}
\end{equation}
where $\alpha(U) = 2(m_\mathrm{e}/M)U^2 N Q^{\mathrm{el}}(U) k_\mathrm{B} T$ arises from the elastic collision operator.

The classification is determined by the determinant $\det(\mathbf{M}) = AC - B^2$. Here, the terms involving the squared electric field cancel out, and the determinant simplifies to
\begin{align}
    \det(\mathbf{M}) = \frac{U Q_\Sigma(U)}{3N} \left( 2\frac{m_\mathrm{e}}{M}U^2 N Q^{\mathrm{el}}(U) k_\mathrm{B} T \right).
\end{align}

The physical quantities ($N, Q_\Sigma(U), m_\mathrm{e}, T$) are strictly positive, thus for all $U > 0$, it results in $\det(\mathbf{M}) > 0$. Therefore, the equation is elliptic in the interior of the domain. As $U \to 0$, the determinant generally vanishes. Consequently, the ellipticity condition fails at the lower energy boundary, classifying the problem as a degenerate elliptic partial differential equation.

It is worth noting that applying the cold gas approximation ($T=0$) makes $\alpha(U)$ vanish, resulting in $\det(\mathbf{M}) = 0$ globally and reclassifying the system as a parabolic equation. However, the mixed derivative terms remain. Consequently, dropping the temperature term does not necessarily accelerate PINN training, as the optimization landscape is primarily hindered by these mixed derivatives rather than pure diffusion. In contrast, including the temperature term ($T>0$) in traditional discretization-based methods often requires the equation to be solved iteratively.

\subsection{Ablation analysis of the ANNet architecture}
\label{appendix:ablation_study}
\begin{figure*}[ht]
\centering
\includegraphics[width=0.9\textwidth]{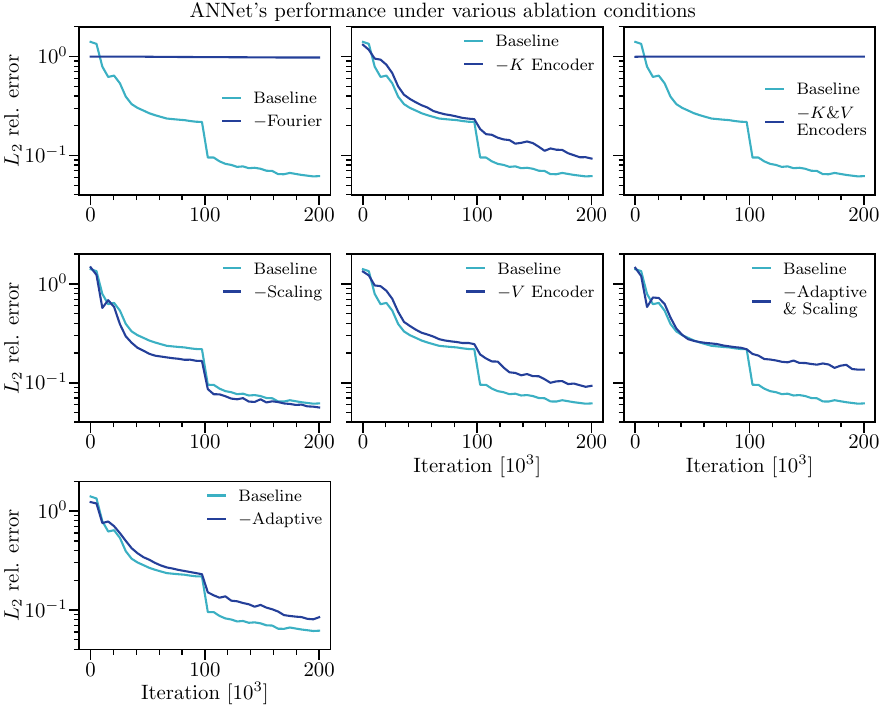}
\caption{\label{fig:ablation_study_v2}Ablation study on the ANNet architecture.}
\end{figure*}

Figure~\ref{fig:ablation_study_v2} presents an ablation study of the ANNet architecture, isolating the effects of the Fourier-feature layer, adaptive activation functions, scaling factor ($1/\sqrt{m}$) in Eq.~\eqref{eq:jannet_attention}, and the gating's $K$ and $V$ encoders. Using the hyperparameter set in Table~\ref{table:hyperparameters}, the results show a clear contrast in how each component influences the model. Removing the Fourier-feature layer or both the $K$ and $V$ encoders causes catastrophic failure, resulting in incorrect (near-trivial) solutions. This strongly indicates that these components are fundamental prerequisites for learning the underlying physics of this problem.

Conversely, removing the $K$ or $V$ encoders yields a functional but less accurate model. These configurations successfully capture the solution's structure and demonstrate converging loss curves, although they are consistently behind the baseline (a case with the complete ANNet architecture) in final accuracy. This suggests that having both encoders active is not strictly necessary for convergence. At the same time, they are essential for optimal performance.

The ablation study further highlights the interplay between the adaptive activation functions and the scaling factor. Removing both of these significantly worsens the convergence rate. Interestingly, removing only the scaling factor does not appear to negatively affect the convergence; in fact, it slightly improves it. For longer training sessions, however, including both the adaptive activation functions and the scaling factor is recommended to ensure robust training across the investigated gas types. Note that with regards to scaling in Eq.~\eqref{eq:jannet_attention}, the learnable parameter $a$ inherently acts as an adaptive scaling factor. Consequently, even if $1/\sqrt{m}$ is omitted, the argument to the activation function remains appropriately scaled by $a$, which effectively helps maintain the argument's variance at an optimal value throughout training.

\subsection{Comparative performance evaluation of ANNet on benchmark PDEs}
\label{appendix:validation}
The ANNet architecture was also validated on two benchmark problems namely the Klein-Gordon and Helmholtz equations, and its performance was quantified against that of the other architectures (FCNNet, WangNet, and ModNet). 

The Klein-Gordon equation is a time-dependent non-linear equation solved here as an initial-boundary value problem. In one spatial dimension, the equation can be written as
\begin{alignat}{2}
  \frac{\partial^2 u}{\partial t^2} + \alpha \frac{\partial^2 u}{\partial x^2} + \beta u + \gamma u^k = f(x,t), &\qquad& (x,t) \in \Omega \times (0,T] \label{eq:klein_gordon}
\end{alignat}
subject to the initial and boundary conditions
\begin{alignat}{2}
  & u(x,0) = g_1(x), &\qquad& x \in \Omega \\
  & \frac{\partial u}{\partial t}(x,0) = g_2(x), &\qquad& x \in \Omega \\
  & u(x,t) = h(x,t), &\qquad& (x,t) \in \partial\Omega \times (0,T]
\end{alignat}
where $\alpha$, $\beta$, $\gamma$ and $k$ are the equation parameters. Here, $f(x,t)$, $g_1(x)=x$, $g_2(x)=0$, and $h(x,t)$ are known functions, and $u(x,t)$ is the target solution. For this study, the spatial domain is defined as the open interval $\Omega = (0,1)$ and the final time is $T=1$, thus the spatiotemporal domain for the PDE is given by $\Omega \times (0,T]$. The parameters are set to $\alpha=-1,\, \beta=0,\, \gamma=1$, and $k=3$. An analytical solution to this problem is used as a reference taking the form
\begin{equation}
\label{eq:klein_gordon_refsol}
u(x,t)=x \cos(5\pi t) + (xt)^3.    
\end{equation}
Accordingly, the forcing term $f(x,t)$ can be derived using Eq.~\eqref{eq:klein_gordon}
\begin{equation}
\begin{aligned}
    f(x,t) = &-25\pi^2x \cos(5\pi t) + 6x^3t + 6\alpha x t^3 \\
             &+\beta(x\cos(5\pi t) + x^3t^3) + \gamma (x \cos(5\pi t) + x^3 t^3)^k.
\end{aligned}
\end{equation}

The second PDE investigated in this section is the Helmholtz equation, which is an elliptic equation. In two dimensions, the equation takes the form
\begin{alignat}{2}
  & \nabla^2 u(x,y) + k^2 u(x,y) = q(x,y), &\qquad& (x,y) \in \Omega = (-1, 1)^2 \label{eq:helmholtz}
\end{alignat}
subject to a homogeneous Dirichlet-type boundary condition
\begin{alignat}{2}
      & u(x,y) = 0, &\qquad& (x,y) \in \partial\Omega
\end{alignat}
Similarly, an analytical function is used as a reference solution to this problem. It is defined as
\begin{equation}
\label{eq:helmholtz_refsol}
    u(x,y) = \sin(a \pi x) \sin (b\pi y).
\end{equation}
The source term for this problem can be derived accordingly
\begin{align}
\label{eq:source}
q(x,y) = \left[k^2 -(a\pi)^2 - (b\pi)^2\right]\sin(a\pi x) \sin(b\pi y),  
\end{align}
where $a=1$, $b=4$, and $k=1$.

The two PDEs were solved using the four architectures specified in Table~\ref{table:architectures} using the following hyperparameters: $m=100,\, L=4,\, n_\mathrm{f}=4000$,  and boundary sampling points of $=512\times 4$. The hyperparameter $\sigma=10$ is used exclusively for ANNet and ModNet. The same exponential learning rate schedule is used; it decays from $10^{-3}$ to $10^{-4}$ until the end of training at $100 \times 10^3$ iterations. Even loss term weighting is used for all loss terms in all cases. In order to see the impact of the choice of the loss function, the models are trained using $L_1$ loss and Mean Squared Error (MSE) loss, which are the typical choices for training PINNs.

The training results are presented in Table~\ref{table:rel_l2_pdes}. Notably, FCNNet fails to converge when using $L_1$ loss, whereas the other architectures, that feature gating mechanisms, demonstrate greater resilience to the choice of loss function. In the MSE loss regime, ModNet (the improved WangNet architecture) demonstrates superior accuracy for both benchmark cases, marking a significant improvement over WangNet and outperforming ANNet. In $L_1$ loss regime, however, ANNet exhibits the highest overall performance, outperforming the other three architectures across both loss regimes. This finding serves as a reference guide when applying ANNet to other complex PDEs.

\begin{figure*}[ht]
    \centering
    \begin{subfigure}[b]{0.9\textwidth}
        \caption{Klein-Gordon equation}
        \includegraphics[width=\linewidth]{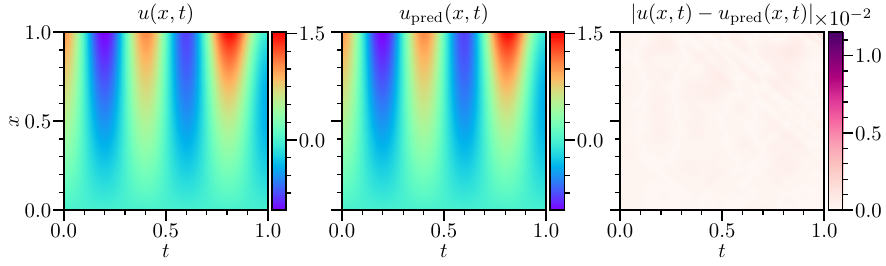}
        \label{fig:center}
    \end{subfigure}

    \begin{subfigure}[b]{0.9\textwidth}
        \caption{Helmholtz equation}
        \includegraphics[width=\linewidth]{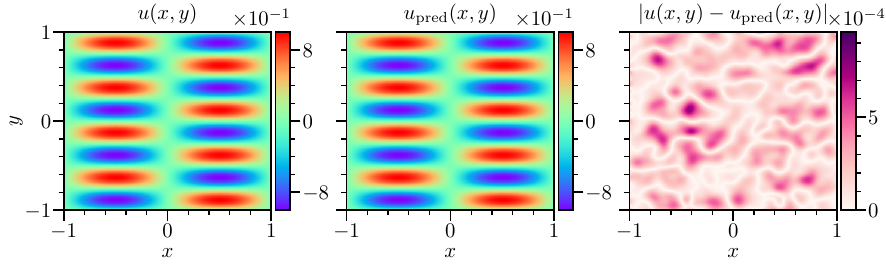}
        \label{fig:bottom}
    \end{subfigure}
    
    \caption{ANNet's solutions $u_\mathrm{pred}$ compared against reference solutions $u$ for (a) the Klein-Gordon equation in $(x,t)$ space and the Helmholtz equation in $(x,y)$ space. The rightmost panels illustrate the absolute error maps.}
    \label{fig:benchmark_pdes}
\end{figure*}

Additionally, Fig.~\ref{fig:benchmark_pdes} compares ANNet's solutions to the two benchmark PDEs, obtained by using $L_1$ loss, against the reference solutions from Eqs.~\eqref{eq:klein_gordon_refsol} and \eqref{eq:helmholtz_refsol}. It shows that the absolute errors are effectively negligible in both cases. It is also worth noting that all networks with a gating mechanism generally perform better with $L_1$ loss compared to their counterparts with MSE loss.

\begin{table}
\centering
\caption{\label{table:rel_l2_pdes} Impact of the loss function choice ($L_1$ vs. MSE) on the accuracy of the investigated PINN architectures. Values represent the $L_2$ relative error $\varepsilon_{L_2}$ ($\times 10^{-2}$) for the Klein-Gordon and Helmholtz benchmark problems. Best scores are presented in bold.}
\begin{ruledtabular}
\begin{tabular}{llllllllll}
    & \multicolumn{2}{c}{Klein-Gordon}  &  \multicolumn{2}{c}{Helmholtz}\\
\hline
 Architecture  & $L_1$& MSE & $L_1$ & MSE \\ 
\hline
FCNNet    & 0.310 & 4.782 & 100.0 & 1.043\\
WangNet   & 0.262 & 1.843 & 0.169 & 0.512\\
ModNet    & 0.050 & \textbf{0.269} & 0.027 & \textbf{0.094}\\
ANNet    & \textbf{0.024} & 0.750 & \textbf{0.023} & 0.106\\
\end{tabular}
\end{ruledtabular}
\end{table}

\subsection{Impact of different \texorpdfstring{$\sigma$}{TEXT} values on ANNet and ModNet}
\label{appendix:varying_sigma}
\begin{figure*}[ht]
\centering
\includegraphics[width=0.8\textwidth]{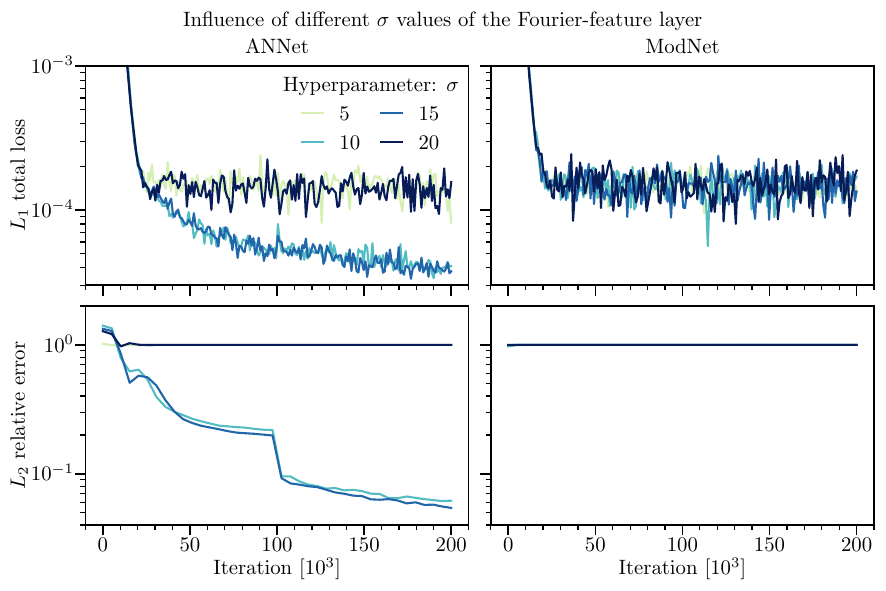}
\caption{\label{fig:diff_sigma}Evolution of $L_1$ total loss and $L_2$ relative error for ANNet and ModNet under varying $\sigma$ values.}
\end{figure*}

A comparative analysis of the training dynamics for ANNet and ModNet is presented in Fig.~\ref{fig:diff_sigma}, specifically examining the influence of the $\sigma$ hyperparameter of the Fourier-feature layer. Based on the hyperparameters in Table~\ref{table:hyperparameters} for the standard argon case, ANNet proves robust within the range of $\sigma \in [10,15]$. In contrast, ModNet demonstrates a complete failure to converge across all tested $\sigma$ values. Thus, ANNet is shown to be effective when properly tuned, while ModNet appears unstable and unsuitable for this specific task.

\section*{References}
\bibliography{main}

@incollection{Loffhagen-2022-ID6521,
author = {Loffhagen, Detlef},
title = {Multiterm and non-local electron {B}oltzmann equation},
booktitle = {Plasma Modeling (Second Edition)},
publisher = {IOP Publishing},
year = {2022},
series = {2053-2563},
type = {Book Chapter},
pages = {3-1 to 3-28},
doi = {10.1088/978-0-7503-3559-1ch3},
isbn = {978-0-7503-3559-1}
}

@article{Winkler-2002-ID1758,
author    = {Winkler, R. and Loffhagen, D. and Sigeneger, F.},
title     = {Temporal and spatial relaxation of electrons in low temperature plasmas},
journal   = {Appl. Surf. Sci.},
year      = {2002},
volume    = {192},
pages     = {50--71},
number    = {1-4},
doi       = {10.1016/S0169-4332(02)00020-X},
month     = {1},
}

@book{Colonna-2022-ID6107,
author    = {Colonna, Gianpiero and D’Angola, Antonio},
title     = {{P}lasma {M}odeling ({S}econd {E}dition)},
publisher = {IOP Publishing},
year      = {2022},
series    = {2053-2563},
isbn      = {978-0-7503-3559-1},
doi       = {10.1088/978-0-7503-3559-1},
month     = {1},
}

@article{Becker-2013-ID3118,
author    = {Becker, M. M. and Loffhagen, D.},
title     = {{D}erivation of {M}oment {E}quations for the {T}heoretical {D}escription of {E}lectrons in {N}onthermal {P}lasmas},
journal   = {Adv. Pure Math.},
year      = {2013},
volume    = {3},
pages     = {343--352},
number    = {3},
doi       = {10.4236/apm.2013.33049},
month     = {5}
}

@book{golant1980fundamentals,
  title={{F}undamentals of {P}lasma {P}hysics},
  author={Golant, V.E. and Zhilinskii, A.P. and Sakharov, I.E.},
  isbn={9780471045939},
  year={1980},
  publisher={Wiley},
  address={New York}
}

@article{Becker-2017-ID4159,
author    = {Becker, M. M. and Kählert, H. and Sun, A. and Bonitz, M. and Loffhagen, D.},
title     = {Advanced fluid modeling and {PIC}/{MCC} simulations of low-pressure ccrf
            discharges},
journal   = {Plasma Sources Sci. Technol. },
year      = {2017},
volume    = {26},
pages     = {044001},
number    = {4},
doi       = {10.1088/1361-6595/aa5cce},
month     = {4}
}

@article{Jovanovic-2023-ID6156,
author    = {Jovanović, Aleksandar P and Loffhagen, Detlef and Becker, Markus M},
title     = {Introduction and verification of {FEDM}, an open-source {FE}ni{CS}-based
            discharge modelling code},
journal   = {Plasma Sources Sci. Technol. },
year      = {2023},
volume    = {32},
pages     = {044003},
number    = {4},
doi       = {10.1088/1361-6595/acc54b},
publisher = {IOP Publishing},
month     = {1},
}

@inproceedings{Cho-2014-ID6453,
author    = {Cho, Kyunghyun and van Merrienboer, Bart and Gulcehre, Caglar and Bahdanau, Dzmitry and Bougares, Fethi and Schwenk, Holger and Bengio, Yoshua},
title     = {{L}earning {P}hrase {R}epresentations using {RNN} {E}ncoder–{D}ecoder for {S}tatistical {M}achine {T}ranslation},
booktitle = {Proceedings of the 2014 Conference on Empirical Methods in Natural Language
            Processing ({EMNLP})},
year      = {2014},
pages     = {1724--1734},
doi       = {10.3115/v1/D14-1179},
month     = {1}}

@article{Raissi-2020-ID6452,
author    = {Raissi, Maziar and Yazdani, Alireza and Karniadakis, George Em},
title     = {Hidden fluid mechanics: {L}earning velocity and pressure fields from flow visualizations},
journal   = {Science},
year      = {2020},
volume    = {367},
pages     = {1026--1030},
number    = {6481},
doi       = {10.1126/science.aaw4741},
month     = {2}}

@article{Sigeneger-1995-ID6451,
author    = {Sigeneger, F and Winkler, R},
title     = {Response of plasma electrons to a spatially embedded electric field impulse.},
journal   = {Phys. Rev. E},
year      = {1995},
volume    = {52},
pages     = {3281--3284},
number    = {3},
doi       = {10.1103/physreve.52.3281},
month     = {9}}

@article{Huang-2025-ID6450,
author    = {Huang, Ziyu and Dong, Chuanfei and Wang, Liang},
title     = {{M}achine-learning heat flux closure for multi-moment fluid modeling of nonlinear {L}andau damping.},
journal = {Proc. Natl. Acad. Sci. U.S.A.},
year      = {2025},
volume    = {122},
pages     = {e2419073122},
number    = {11},
doi       = {10.1073/pnas.2419073122},
month     = {3}}

@article{Li-2024-ID6449,
author    = {Li, Zongyi and Zheng, Hongkai and Kovachki, Nikola and Jin, David and Chen, Haoxuan and Liu, Burigede and Azizzadenesheli, Kamyar and Anandkumar, Anima},
title     = {{P}hysics-{I}nformed {N}eural {O}perator for {L}earning {P}artial {D}ifferential {E}quations},
journal   = {{ACM} / {IMS} J. Data Sci.},
year      = {2024},
volume    = {1},
pages     = {1--27},
number    = {3},
doi       = {10.1145/3648506},
month     = {1}}

@article{Wang-2021-ID6448,
author    = {Wang, Sifan and Teng, Yujun and Perdikaris, Paris},
title     = {{U}nderstanding and {M}itigating {G}radient {F}low {P}athologies in {P}hysics-{I}nformed {N}eural {N}etworks},
journal   = {{SIAM} J. Sci. Comput.},
year      = {2021},
volume    = {43},
pages     = {A3055--A3081},
number    = {5},
doi       = {10.1137/20M1318043},
month     = {1}}

@article{Hu-2024-ID6447,
author    = {Hu, Zheyuan and Shukla, Khemraj and Karniadakis, George Em and Kawaguchi,
            Kenji},
title     = {{T}ackling the curse of dimensionality with physics-informed neural networks},
journal   = {Neural Netw.},
year      = {2024},
volume    = {176},
pages     = {106369},
doi       = {10.1016/j.neunet.2024.106369},
month     = {8}}

@article{Sirignano-2018-ID6446,
author    = {Sirignano, Justin and Spiliopoulos, Konstantinos},
title     = {{DGM}: {A} deep learning algorithm for solving partial differential equations},
journal   = {J. Comput. Phys.},
year      = {2018},
volume    = {375},
pages     = {1339--1364},
doi       = {10.1016/j.jcp.2018.08.029},
month     = {12}}

@article{Jumper-2021-ID6444,
author    = {Jumper, John and Evans, Richard and Pritzel, Alexander and Green, Tim and Figurnov, Michael and Ronneberger, Olaf and Tunyasuvunakool, Kathryn and Bates, Russ and Žídek, Augustin and Potapenko, Anna and Bridgland, Alex and Meyer, Clemens and Kohl, Simon A A and Ballard, Andrew J and Cowie, Andrew and Romera-Paredes, Bernardino and Nikolov, Stanislav and Jain, Rishub and Adler, Jonas and Back, Trevor and Petersen, Stig and Reiman, David and Clancy, Ellen and Zielinski, Michal and Steinegger, Martin and Pacholska, Michalina and Berghammer, Tamas and Bodenstein, Sebastian and Silver, David and Vinyals, Oriol and Senior, Andrew W and Kavukcuoglu, Koray and Kohli, Pushmeet and Hassabis, Demis},
title     = {{H}ighly accurate protein structure prediction with {A}lpha{F}old.},
journal   = {Nature},
year      = {2021},
volume    = {596},
pages     = {583--589},
number    = {7873},
doi       = {10.1038/s41586-021-03819-2},
month     = {8}}

@techreport{Baker-2019-ID6443,
author    = {Baker, Nathan and Alexander, Frank and Bremer, Timo and Hagberg, Aric and Kevrekidis, Yannis and Najm, Habib and Parashar, Manish and Patra, Abani and Sethian, James and Wild, Stefan and Willcox, Karen and Lee, Steven},
title     = {{W}orkshop {R}eport on {B}asic {R}esearch {N}eeds for {S}cientific {M}achine {L}earning: {C}ore {T}echnologies for {A}rtificial {I}ntelligence},
year      = {2019},
institution = {USDOE Office of Science (SC) (United States)},
doi       = {10.2172/1478744},
month     = {1}}

@article{Kim-2023-ID6440,
author    = {Kim, Jin Seok and Denpoh, Kazuki and Kawaguchi, Satoru and Satoh, Kohki and
            Matsukuma, Masaaki},
title     = {{N}umerical {s}trategy for solving the {B}oltzmann equation with variable {E}/{N} using physics-informed neural networks},
journal   = {J. Phys. D: Appl. Phys.},
year      = {2023},
volume    = {56},
pages     = {344002},
number    = {34},
doi       = {10.1088/1361-6463/accbcf},
month     = {1}}

@article{Kawaguchi-2020-ID6439,
author    = {Kawaguchi, S. and Takahashi, K. and Ohkama, H. and Satoh, K.},
title     = {{D}eep learning for solving the {B}oltzmann equation of electrons in weakly ionized plasma},
journal   = {Plasma Sources Sci. Technol.},
year      = {2020},
volume    = {29},
pages     = {025021},
doi       = {10.1088/1361-6595/ab6074},
month     = {2}}

@article{Kortshagen-1993-ID6438,
author    = {Kortshagen, U.},
title     = {{O}n the influence of energy transfer efficiency on the electron energy distribution function in {HF} sustained rare gas plasmas: experimental and numerical study},
journal   = {J. Phys. D: Appl. Phys.},
year      = {1993},
volume    = {26},
pages     = {1230--1238},
doi       = {10.1088/0022-3727/26/8/012},
month     = {8}}

@article{Winkler-1987-ID6437,
author    = {Winkler, R. and Dilonardo, M. and Capitelli, M. and Wilhelm, J.},
title     = {{T}ime-dependent solution of {B}oltzmann equation in rf plasmas: A comparison with the effective field approximation},
journal   = {Plasma Chem. Plasma Process.},
year      = {1987},
volume    = {7},
pages     = {125--137},
number    = {1},
doi       = {10.1007/BF01016003},
month     = {1}}

@article{Tsendin-0000-ID6436,
author    = {Tsendin, Lev D.},
title     = {{E}nergy distribution of electrons in a weakly ionized current-carrying plasma with a transverse inhomogeneity},
journal   = {Sov. Phys. – {JETP}},
year      = {1974},
volume    = {39},
pages     = {805},
month     = {5}
}

@article{Anirudh-2022-ID6250,
author    = {Anirudh, Rushil and Archibald, Rick and Asif, M. Salman and Becker, Markus M. and Benkadda, Sadruddin and Bremer, Peer-Timo and Budé, Rick H. S. and Chang, C. S. and Chen, Lei and Churchill, R. M. and Citrin, Jonathan and Gaffney, Jim A and Gainaru, Ana and Gekelman, Walter and Gibbs, Tom and Hamaguchi, Satoshi and Hill, Christian and Humbird, Kelli and Jalas, Sören and Kawaguchi, Satoru and Kim, Gon-Ho and Kirchen, Manuel and Klasky, Scott and Kline, John L. and Krushelnick, Karl and Kustowski, Bogdan and Lapenta, Giovanni and Li, Wenting and Ma, Tammy and Mason, Nigel J. and Mesbah, Ali and Michoski, Craig and Munson, Todd and Murakami, Izumi and Najm, Habib N. and Olofsson, K. Erik J. and Park, Seolhye and Peterson, J. Luc and Probst, Michael and Pugmire, Dave and Sammuli, Brian and Sawlani, Kapil and Scheinker, Alexander and Schissel, David P. and Shalloo, Rob J. and Shinagawa, Jun and Seong, Jaegu and Spears, Brian K. and Tennyson, Jonathan and Thiagarajan, Jayaraman and Ticoş, Catalin M. and Trieschmann, Jan and Dijk, Jan van and Essen, Brian Van and Ventzek, Peter and Wang, Haimin and Wang, Jason T. L. and Wang, Zhehui and Wende, Kristian and Xu, Xueqiao and Yamada, Hiroshi and Yokoyama, Tatsuya and Zhang, Xinhua},
title     = {2022 Review of {D}ata-{D}riven {P}lasma {S}cience},
journal   = {{IEEE} Trans. Plasma Sci.},
year      = {2022},
month     = {5},
doi       = {10.1109/TPS.2023.3268170}
}

@article{Hornik-1991-ID6198,
author    = {Hornik, Kurt},
title     = {{A}pproximation capabilities of multilayer feedforward networks},
journal   = {Neural Netw.},
year      = {1991},
volume    = {4},
pages     = {251--257},
number    = {1},
doi       = {10.1016/0893-6080(91)90009-T},
month     = {1}}

@article{Hornik-1989-ID6197,
author    = {Hornik, Kurt and Stinchcombe, Maxwell and White, Halbert},
title     = {{M}ultilayer feedforward networks are universal approximators},
journal   = {Neural Netw.},
year      = {1989},
volume    = {2},
pages     = {359--366},
number    = {5},
doi       = {10.1016/0893-6080(89)90020-8},
month     = {1}}

@article{Wang-2021-ID6196,
author    = {Wang, Sifan and Wang, Hanwen and Perdikaris, Paris},
title     = {{L}earning the solution operator of parametric partial differential equations with physics-informed Deep{ON}ets.},
journal   = {Sci. Adv.},
year      = {2021},
volume    = {7},
pages     = {eabi8605},
number    = {40},
doi       = {10.1126/sciadv.abi8605},
month     = {10},
}

@article{Lu-2021-ID6189,
author    = {Lu, Lu and Meng, Xuhui and Mao, Zhiping and Karniadakis, George Em},
title     = {Deep{XDE}: {A} {D}eep {L}earning {L}ibrary for {S}olving {D}ifferential {E}quations},
journal   = {{SIAM} Rev. },
year      = {2021},
volume    = {63},
pages     = {208--228},
number    = {1},
doi       = {10.1137/19M1274067},
month     = {1}}

@article{Raissi-2019-ID6187,
author    = {Raissi, M and Perdikaris, P and Karniadakis, G E},
title     = {{P}hysics-informed neural networks: {A} deep learning framework for solving forward and inverse problems involving nonlinear partial differential equations},
journal   = {J. Comput. Phys.},
year      = {2019},
volume    = {378},
pages     = {686--707},
doi       = {10.1016/j.jcp.2018.10.045},
month     = {2}}

@article{Lagaris-1998-ID6186,
author    = {Lagaris, I E and Likas, A and Fotiadis, D I},
title     = {{A}rtificial neural networks for solving ordinary and partial differential equations.},
journal   = {{IEEE} Trans. Neural Netw. },
year      = {1998},
volume    = {9},
pages     = {987--1000},
number    = {5},
doi       = {10.1109/72.712178},
month     = {1}}

@article{Kawaguchi-2022-ID6158,
author    = {Kawaguchi, Satoru and Murakami, Tomoyuki},
title     = {{P}hysics-informed neural networks for solving the {B}oltzmann equation of the electron velocity distribution function in weakly ionized plasmas},
journal   = {Jpn. J. Appl. Phys.},
year      = {2022},
volume    = {61},
pages     = {086002},
number    = {8},
doi       = {10.35848/1347-4065/ac7afb},
month     = {8}
}

@article{Adamovich-2022-ID6011,
author    = {Adamovich, I and Agarwal, S and Ahedo, E and Alves, L L and Baalrud, S and
            Babaeva, N and Bogaerts, A and Bourdon, A and Bruggeman, P J and Canal, C
            and Choi, E H and Coulombe, S and Donkó, Z and Graves, D B and Hamaguchi,
            S and Hegemann, D and Hori, M and Kim, H-H and Kroesen, G M W and Kushner,
            M J and Laricchiuta, A and Li, X and Magin, T E and Mededovic Thagard, S
            and Miller, V and Murphy, A B and Oehrlein, G S and Puac, N and Sankaran, R
            M and Samukawa, S and Shiratani, M and Šimek, M and Tarasenko, N and
            Terashima, K and Thomas Jr, E and Trieschmann, J and Tsikata, S and Turner,
            M M and van der Walt, I J and van de Sanden, M C M and von Woedtke, T},
title     = {{T}he 2022 {P}lasma {R}oadmap: {L}ow temperature plasma science and technology},
journal   = {J. Phys. D: Appl. Phys.},
year      = {2022},
volume    = {55},
pages     = {373001},
number    = {37},
doi       = {10.1088/1361-6463/ac5e1c},
month     = {1},
}

@article{Rapp-1965-ID5358,
author    = {Rapp, Donald and Englander‐Golden, Paula},
title     = {{T}otal {C}ross Sections for {I}onization and {A}ttachment in {G}ases by {E}lectron {I}mpact. {I}. {P}ositive {I}onization},
journal   = {J. Chem. Phys.},
year      = {1965},
volume    = {43},
pages     = {1464--1479},
number    = {5},
doi       = {10.1063/1.1696957},
month     = {1},
}

@article{Adamovich-2017-ID4238,
author    = {Adamovich, I. and Baalrud, S. D. and Bogaerts, A. and Bruggeman, P. J. and
            Cappelli, M. and Colombo, V. and Czarnetzki, U. and Ebert, U. and Eden, J.
            G. and Favia, P. and Graves, D. B. and Hamaguchi, S. and Hieftje, G. and
            Hori, M. and Kaganovich, I. D. and Kortshagen, U. and Kushner, M. J. and
            Mason, N. J. and Mazouffre, S. and Mededovic Thagard, S. and Metelmann,
            H.-R. and Mizuno, A. and Moreau, E. and Murphy, A. B. and Niemira, B. A.
            and Oehrlein, G. S. and Petrovic, Z. Lj and Pitchford, L. C. and Pu, Y.-K.
            and Rauf, S. and Sakai, O. and Samukawa, S. and Starikovskaia, S. and
            Tennyson, J. and Terashima, K. and Turner, M. M. and van de Sanden, M. C.
            M. and Vardelle, A.},
title     = {{T}he 2017 {P}lasma {R}oadmap: {L}ow temperature plasma science and technology},
journal   = {J. Phys. D: Appl. Phys.},
year      = {2017},
volume    = {50},
pages     = {323001},
number    = {32},
doi       = {10.1088/1361-6463/aa76f5},
month     = {8}
}

@article{Loffhagen-2005-ID3906,
author    = {Loffhagen, D.},
title     = {{I}mpact of {E}lectron–{E}lectron {C}ollisions on the {S}patial {E}lectron {R}elaxation in {N}on-{I}sothermal {P}lasmas},
journal   = {Plasma Chem. Plasma Process.},
year      = {2005},
volume    = {25},
pages     = {519--538},
number    = {5},
doi       = {10.1007/s11090-005-4997-y},
month     = {1}
}

@article{Arslanbekov-1998-ID3840,
author    = {Arslanbekov, Robert R. and Kudryavtsev, Anatoly A.},
title     = {{M}odeling of nonlocal electron kinetics in a low-pressure afterglow plasma},
journal   = {Phys. Rev. E},
year      = {1998},
volume    = {58},
pages     = {7785--7798},
doi       = {10.1103/PhysRevE.58.7785},
month     = {12}
}

@article{Kortshagen-1996-ID3034,
author    = {Kortshagen, U. and Busch, C. and Tsendin, L. D.},
title     = {{O}n simplifying approaches to the solution of the {B}oltzmann equation in spatially inhomogeneous plasmas},
journal   = {Plasma Sources Sci. Technol.},
year      = {1996},
volume    = {5},
pages     = {1--17},
doi       = {10.1088/0963-0252/5/1/001},
month     = {2}
}

@article{Godyak-1992-ID3006,
author    = {Godyak, V. A. and Piejak, R. B. and Alexandrovich, B. M.},
title     = {{M}easurement of electron energy distribution in low-pressure {RF} discharges},
journal   = {Plasma Sources Sci. Technol.},
year      = {1992},
volume    = {1},
pages     = {36--58},
doi       = {10.1088/0963-0252/1/1/006},
month     = {3}
}

@article{Hagelaar-2005-ID2276,
author    = {Hagelaar, G. J. M. and Pitchford, L. C.},
title     = {{S}olving the {B}oltzmann equation to obtain electron transport coefficients and rate coefficients for fluid models},
journal   = {Plasma Sources Sci. Technol.},
year      = {2005},
volume    = {14},
pages     = {722--733},
doi       = {10.1088/0963-0252/14/4/011},
month     = {11}
}

@article{Loffhagen-2002-ID1724,
author    = {Loffhagen, D. and Winkler, R. and Donkó, Z.},
title     = {{B}oltzmann equation and {M}onte {C}arlo analysis of the spatiotemporal electron relaxation in nonisothermal plasmas},
journal   = {Eur. Phys. J. {AP}},
year      = {2002},
volume    = {18},
pages     = {189--200},
number    = {3},
doi       = {10.1051/epjap:2002040},
month     = {1}
}

@article{Sigeneger-1999-ID1309,
author    = {Sigeneger, F. and Winkler, R.},
title     = {{N}onlocal transport and dissipation properties of electrons in inhomogeneous plasmas},
journal   = {{IEEE} Trans. Plasma Sci.},
year      = {1999},
volume    = {27},
pages     = {1254--1261},
doi       = {10.1109/27.799801},
month     = {10}
}

@article{Sigeneger-1998-ID1242,
author    = {Sigeneger, F. and Golubovskii, Yu. B. and Porokhova, I. A. and Winkler, R.},
title     = {{O}n the {N}onlocal {E}lectron {K}inetics in s- and p-{S}triations of {DC} {G}low {D}ischarge {P}lasmas: {I}. {E}lectron {E}stablishment in {S}triation-{l}ike {F}ields},
journal   = {Plasma Chem. Plasma Process.},
year      = {1998},
volume    = {18},
pages     = {153--180},
number    = {2},
doi       = {10.1023/A:1021694231064},
month     = {1}
}

@article{Leyh-1998-ID1222,
author    = {Leyh, H. and Loffhagen, D. and Winkler, R.},
title     = {{A} new multi-term solution technique for the electron {B}oltzmann equation of weakly ionized steady-state plasmas},
journal   = {Comput. Phys. Commun.},
year      = {1998},
volume    = {113},
pages     = {33--48},
number    = {1},
doi       = {10.1016/S0010-4655(98)00062-9},
month     = {9}
}

@article{Winkler-1997-ID1137,
author    = {Winkler, R. and Petrov, G. and Sigeneger, F. and Uhrlandt, D.},
title     = {{S}trict calculation of electron energy distribution functions in inhomogeneous plasmas},
journal   = {Plasma Sources Sci. Technol.},
year      = {1997},
volume    = {6},
pages     = {118--132},
number    = {2},
doi       = {10.1088/0963-0252/6/2/005},
month     = {5}
}

@article{Sigeneger-1996-ID1050,
author    = {Sigeneger, F. and Winkler, R.},
title     = {Response of the {E}lectron {K}inetics on {S}patial {D}isturbances of the {E}lectric {F}ield in {N}onisothermal {P}lasmas},
journal   = {Contrib. Plasma Phys.},
year      = {1996},
volume    = {36},
pages     = {551--571},
number    = {5},
doi       = {10.1002/ctpp.2150360503},
month     = {1}
}

@article{Tsendin-1995-ID976,
author    = {Tsendin, L D},
title     = {{E}lectron kinetics in non-uniform glow discharge plasmas},
journal   = {Plasma Sources Sci. Technol.},
year      = {1995},
volume    = {4},
pages     = {200--211},
number    = {2},
doi       = {10.1088/0963-0252/4/2/004},
month     = {1}
}

@incollection{Hayashi-1992-ID751,
  author    = {Hayashi, M.},
  title     = {Electron collision cross sections},
  booktitle = {Plasma {M}aterial {S}cience {H}andbook},
  editor    = {{Japan Society for the Promotion of Science}},
  publisher = {Ohmsha, Ltd},
  address   = {Tokyo},
  year      = {1992},
  pages     = {748--766}
}

@article{Fon-1981-ID6454,
author    = {Fon, W. C. and Berrington, K. A. and Hibbert, A.},
title     = {{T}he elastic scattering of electrons from inert gases. {I}. {H}elium},
journal   = {J. Phys. B: At. Mol. Phys.},
year      = {1981},
volume    = {14},
pages     = {307--321},
doi       = {10.1088/0022-3700/14/2/014},
month     = {1}}

@article{Mitroy-1990-ID6455,
author    = {Mitroy, Jim},
title     = {{T}he momentum transfer cross section for krypton},
journal   = {Aust. J. Phys.},
year      = {1990},
volume    = {43},
pages     = {19},
doi       = {10.1071/PH900019},
month     = {1}}

@article{Specht-1980-ID6456,
author    = {Specht, L. T. and Lawton, S. A. and DeTemple, T. A.},
title     = {{E}lectron ionization and excitation coefficients for argon, krypton, and xenon in the low {E}/{N} region},
journal   = {J. Appl. Phys.},
year      = {1980},
volume    = {51},
pages     = {166--170},
doi       = {10.1063/1.327395},
month     = {1}}

@article{deHeer-1979-ID6457,
author    = {de Heer, F. J. and Jansen, R. H. J. and van der Kaay, W.},
title     = {{T}otal cross sections for electron scattering by {N}e, {A}r, {K}r and {X}e},
journal   = {J. Phys. B: At. Mol. Phys.},
year      = {1979},
volume    = {12},
pages     = {979--1002},
doi       = {10.1088/0022-3700/12/6/016},
month     = {3}}

@article{Wetzel-1987-ID6458,
author    = {Wetzel, RC and Baiocchi, FA and Hayes, TR and Freund, RS},
title     = {{A}bsolute cross sections for electron-impact ionization of the rare-gas atoms by the fast-neutral-beam method.},
journal   = {Phys. Rev. A},
year      = {1987},
volume    = {35},
pages     = {559--577},
number    = {2},
doi       = {10.1103/physreva.35.559},
month     = {1}}

@article{Lam-1982-ID6459,
author    = {Sin Fai Lam, L. T.},
title     = {{R}elativistic effects in electron scattering by atoms. {III}. {E}lastic scattering by krypton, xenon and radon},
journal   = {J. Phys. B: At. Mol. Phys.},
year      = {1982},
volume    = {15},
pages     = {119--142},
doi       = {10.1088/0022-3700/15/1/020},
month     = {1}}

@article{Puech-1991-ID6460,
author    = {Puech, V. and Mizzi, S.},
title     = {{C}ollision cross sections and transport parameters in neon and xenon},
journal = {J. Phys. D: Appl. Phys.},
year      = {1991},
volume    = {24},
pages     = {1974--1985},
doi       = {10.1088/0022-3727/24/11/011},
month     = {11}}

@article{Gergs-2023-ID6462,
author    = {Gergs, Tobias and Mussenbrock, Thomas and Trieschmann, Jan},
title     = {{P}hysics-separating artificial neural networks for predicting sputtering and thin film deposition of {A}l{N} in {A}r/{N}2 discharges on experimental timescales},
journal = {J. Phys. D: Appl. Phys.},
year      = {2023},
volume    = {56},
pages     = {194001},
number    = {19},
doi       = {10.1088/1361-6463/acc07e},
month     = {1}}

@article{Trieschmann-2023-ID6461,
author    = {Trieschmann, Jan and Vialetto, Luca and Gergs, Tobias},
title     = {{R}eview: {M}achine learning for advancing low-temperature plasma modeling and simulation},
journal = {J. Micro/Nanopattern. Mats. Metro.},
year      = {2023},
volume    = {22},
number    = {04},
doi       = {10.1117/1.JMM.22.4.041504},
month     = {1}}

@article{Baydin2017,
author = {Baydin, At\i{}l\i{}m G\"{u}nes and Pearlmutter, Barak A. and Radul, Alexey Andreyevich and Siskind, Jeffrey Mark},
title = {{A}utomatic differentiation in machine learning: a survey},
year = {2017},
issue_date = {January 2017},
publisher = {JMLR.org},
volume = {18},
number = {1},
issn = {1532-4435},
journal = {J. Mach. Learn. Res.},
month = jan,
pages = {5595–5637},
numpages = {43}
}

@inproceedings{Cranmer2020,
author = {Cranmer, Miles and Sanchez-Gonzalez, Alvaro and Battaglia, Peter and Xu, Rui and Cranmer, Kyle and Spergel, David and Ho, Shirley},
title = {{D}iscovering symbolic models from deep learning with inductive biases},
year = {2020},
isbn = {9781713829546},
publisher = {Curran Associates Inc.},
address = {Red Hook, NY, USA},
booktitle = {Proceedings of the 34th International Conference on Neural Information Processing Systems},
articleno = {1462},
numpages = {14},
location = {Vancouver, BC, Canada},
series = {NIPS '20}
}

@inproceedings{Krishnapriyan2021,
author = {Krishnapriyan, Aditi S. and Gholami, Amir and Zhe, Shandian and Kirby, Robert M. and Mahoney, Michael W.},
title = {Characterizing possible failure modes in physics-informed neural networks},
year = {2021},
isbn = {9781713845393},
publisher = {Curran Associates Inc.},
address = {Red Hook, NY, USA},
booktitle = {Proceedings of the 35th International Conference on Neural Information Processing Systems},
articleno = {2033},
numpages = {13},
series = {NIPS '21}
}

@InProceedings{Rahaman2019,
  title = 	 {{O}n the {S}pectral {B}ias of {N}eural {N}etworks},
  author =       {Rahaman, Nasim and Baratin, Aristide and Arpit, Devansh and Draxler, Felix and Lin, Min and Hamprecht, Fred and Bengio, Yoshua and Courville, Aaron},
  booktitle = 	 {Proceedings of the 36th International Conference on Machine Learning},
  pages = 	 {5301--5310},
  year = 	 {2019},
  volume = 	 {97},
  series = 	 {Proceedings of Machine Learning Research},
  month = 	 {09--15 Jun},
  publisher =    {PMLR}
}

@inproceedings{Tancik2020,
author = {Tancik, Matthew and Srinivasan, Pratul P. and Mildenhall, Ben and Fridovich-Keil, Sara and Raghavan, Nithin and Singhal, Utkarsh and Ramamoorthi, Ravi and Barron, Jonathan T. and Ng, Ren},
title = {{F}ourier features let networks learn high frequency functions in low dimensional domains},
year = {2020},
isbn = {9781713829546},
publisher = {Curran Associates Inc.},
address = {Red Hook, NY, USA},
articleno = {632},
numpages = {11},
location = {Vancouver, BC, Canada},
series = {NIPS '20},
booktitle = {Proceedings of the 34th International Conference on Neural Information Processing Systems},
}

@inproceedings{Daw2023,
author = {Daw, Arka and Bu, Jie and Wang, Sifan and Perdikaris, Paris and Karpatne, Anuj},
title = {{M}itigating propagation failures in physics-informed neural networks using retain-resample-release (r3) sampling},
year = {2023},
publisher = {JMLR.org},
articleno = {288},
numpages = {39},
location = {Honolulu, Hawaii, USA},
series = {ICML'23},
booktitle = {Proceedings of the 40th International Conference on Machine Learning},
}

@misc{AlAradi2018,
      title={{S}olving {N}onlinear and {H}igh-{D}imensional {P}artial {D}ifferential {E}quations via {D}eep {L}earning}, 
      author={Ali Al-Aradi and Adolfo Correia and Danilo Naiff and Gabriel Jardim and Yuri Saporito},
      year={2018},
      eprint={1811.08782},
      archivePrefix={arXiv},
      primaryClass={q-fin.CP},
      url={https://arxiv.org/abs/1811.08782}, 
}

@inproceedings{Vaswani2017,
author = {Vaswani, Ashish and Shazeer, Noam and Parmar, Niki and Uszkoreit, Jakob and Jones, Llion and Gomez, Aidan N. and Kaiser, \L{}ukasz and Polosukhin, Illia},
title = {Attention is all you need},
year = {2017},
isbn = {9781510860964},
publisher = {Curran Associates Inc.},
address = {Red Hook, NY, USA},
booktitle = {Proceedings of the 31st International Conference on Neural Information Processing Systems},
pages = {6000–6010},
numpages = {11},
location = {Long Beach, California, USA},
series = {NIPS'17}
}

@inproceedings{Glorot2010,
title =  {{U}nderstanding the difficulty of training deep feedforward neural networks},
author =  {Glorot, Xavier and Bengio, Yoshua},
booktitle =  {Proceedings of the Thirteenth International Conference on Artificial Intelligence and Statistics},
pages =  {249--256},
year =  {2010},
volume =  {9},
series =  {Proceedings of Machine Learning Research},
address =  {Chia Laguna Resort, Sardinia, Italy},
month =  {05},
publisher = {PMLR}
}

@inproceedings{Kingma2015,
  author    = {Kingma, Diederik P. and Ba, Jimmy},
  title     = {{A}dam: {A} Method for {S}tochastic {O}ptimization},
  booktitle = {International Conference on Learning Representations (ICLR)},
  year      = {2015},
  url       = {https://arxiv.org/abs/1412.6980}
}

@inbook{Paszke2019,
author = {Paszke, Adam and Gross, Sam and Massa, Francisco and Lerer, Adam and Bradbury, James and Chanan, Gregory and Killeen, Trevor and Lin, Zeming and Gimelshein, Natalia and Antiga, Luca and Desmaison, Alban and K\"{o}pf, Andreas and Yang, Edward and DeVito, Zach and Raison, Martin and Tejani, Alykhan and Chilamkurthy, Sasank and Steiner, Benoit and Fang, Lu and Bai, Junjie and Chintala, Soumith},
title = {{P}y{T}orch: an imperative style, high-performance deep learning library},
year = {2019},
publisher = {Curran Associates Inc.},
address = {Red Hook, NY, USA},
booktitle = {Proceedings of the 33rd International Conference on Neural Information Processing Systems},
articleno = {721},
numpages = {12}
}

@inbook{Hayashi2003,
    author = {Hayashi, M},
    title = {{B}ibliography of electron and photon cross sections with atoms and molecules published in the 20th century},
    publisher = {report NIFS-DAT-72 of the National Institute for Fusion Science of Japan},
    year = 2003
}

@article{Jagtap2020,
    author = {Jagtap, Ameya D. and Kawaguchi, Kenji and Em Karniadakis, George},
    title = {{L}ocally adaptive activation functions with slope recovery for deep and physics-informed neural networks},
    journal = {Proc. R. Soc. A.},
    volume = {476},
    number = {2239},
    pages = {20200334},
    year = {2020},
    month = {07},
    issn = {1364-5021},
    doi = {10.1098/rspa.2020.0334}
}

@article{Hayashi_1983,
doi = {10.1088/0022-3727/16/4/018},
year = {1983},
month = {4},
volume = {16},
number = {4},
pages = {581},
author = {M Hayashi},
title = {{D}etermination of electron-xenon total excitation cross-sections, from threshold to 100 eV, from experimental values of {T}ownsend's $\alpha$},
journal = {J. Phys. D: Appl. Phys. }
}

@book{raizer1991gas,
  title={Gas Discharge Physics},
  author={Raizer, Y. P.},
  editor={Allen, J. E.},
  year={1991},
  publisher={Springer Berlin Heidelberg},
  isbn={978-3-642-64760-4}
}

@article{Tsendin_2010,
doi = {10.3367/UFNe.0180.201002b.0139},
year = {2010},
month = {may},
publisher = {},
volume = {53},
number = {2},
pages = {133},
author = {Tsendin, Lev D},
title = {Nonlocal electron kinetics in gas-discharge plasma},
journal = {Phys.-Usp.}
}

\end{document}